\documentclass[acmsmall,natbib=true,preprint]{acmart-arxiv}
\usepackage[normalem]{ulem}
\usepackage{csquotes}
\usepackage{url}
\usepackage{xcolor} 
\usepackage[ruled]{algorithm2e}
\usepackage{amsmath}
\usepackage{booktabs}
\usepackage{enumerate}
\usepackage[shortlabels]{enumitem}
\usepackage[mathscr]{eucal}
\usepackage{graphicx}
\usepackage{makecell}
\usepackage{mathtools}
\usepackage{multirow}
\usepackage[font=rm]{caption}
\usepackage{subcaption}
\DeclareCaptionType{copyrightbox}
\usepackage{shortvrb}
\usepackage{tabularx}
\usepackage{threeparttable}
\usepackage{verbatim}
\usepackage{xspace}
\usepackage{listings}
\usepackage{tikz}
\usepackage{balance}

\renewcommand{\cite}{\citep}

\newcommand{\combsum}{\mbox{CombSUM}}
\newcommand{\sid}{{\ensuremath{s_{i,d}}}}

\newcommand{\metric}[1]{{\mbox{#1}}}

\newcommand\method[1]{{\sf\small{#1}}}

\newcommand{\opstyle}[1]{\mbox{{\textsc{#1}}}}

\newcommand{\wand}{\opstyle{WAND}\xspace}
\newcommand{\bmw}{\opstyle{BMW}\xspace}
\newcommand{\vbmw}{\opstyle{VBMW}\xspace}

\newcommand{\maxscore}{\opstyle{MaxScore}\xspace}
\newcommand{\taat}{\opstyle{TaaT}\xspace}
\newcommand{\daat}{\opstyle{DaaT}\xspace}

\newcommand{\myparagraph}[1]%
  {\paragraph*{\hspace*{-\parindent}\normalsize\bf#1.}}
\newcommand{\mycaption}[1]{\caption{\normalfont{#1}}}

\newcommand\kb[1]{$#1$\,kiB}

\setcopyright{acmlicensed}

\acmYear{2018}
\acmMonth{11}
\received{November, 2018}

\begin{document}

\title{Boosting Search Performance Using Query Variations}

\newcommand{\ausuper}[1]{\raisebox{1.0ex}{\normalsize\sf{#1}}}

\author{Rodger Benham}
\orcid{0000-0001-6319-5165}
\affiliation{
  \institution{RMIT University}
  \city{Melbourne}
  \state{Australia}
}

\author{Joel Mackenzie}
\orcid{0000-0001-7992-4633}
\affiliation{
  \institution{RMIT University}
  \city{Melbourne}
  \state{Australia}
}

\author{Alistair Moffat}
\orcid{0000-0002-6638-0232}
\affiliation{
  \institution{The University of Melbourne}
  \city{Melbourne}
  \state{Australia}
}

\author{J. Shane Culpepper}
\orcid{0000-0002-1902-9087}
\affiliation{
  \institution{RMIT University}
  \city{Melbourne}
  \state{Australia}
}

\begin{abstract}
Rank fusion is a powerful technique that allows multiple sources of
information to be combined into a single result set.
However, to date fusion has not been regarded as being cost-effective
in cases where strict per-query efficiency guarantees are required,
such as in web search.
In this work we propose a novel solution to rank fusion by splitting
the computation into two parts -- one phase that is carried out
offline to generate pre-computed centroid answers for queries with
broadly similar information needs, and then a second online phase
that uses the corresponding topic centroid to compute a result page
for each query.
We explore efficiency improvements to classic fusion algorithms whose
costs can be amortized as a pre-processing step, and can then be
combined with re-ranking approaches to dramatically improve
effectiveness in multi-stage retrieval systems with little efficiency
overhead at query time.
Experimental results using the ClueWeb12B collection and the UQV100
query variations demonstrate that centroid-based approaches allow
improved retrieval effectiveness at little or no loss in query
throughput or latency, and with reasonable pre-processing
requirements.
We additionally show that queries that do not match any of the
pre-computed clusters can be accurately identified and efficiently
processed in our proposed ranking pipeline.

\end{abstract}

\maketitle

\section{Introduction}\label{sec-intro}

Rank fusion is used to combine knowledge from different result sets
into a single highly-effective answer page.
The fusion can be score-based, in which the retrieval scores of
documents are aggregated; or rank-based, in which documents are
assigned a weighting based solely on their positions in the separate
lists.
In both cases, the new top-$k$ result set is derived by re-sorting
the documents after aggregate weightings are computed.
{\citet{vc99irj}} describe the effects that allow fusion to produce a
more effective response: taking advantage of diversity in document
representation (\emph{skimming}); building consensus among ranked
lists (\emph{chorus}); and catering for differences in quality of
rankers (\emph{dark horse}).
{\citet{bccc93-sigir}} show that a simple unsupervised fusion of all
related queries on the same system yields greater effectiveness than
fusing one query issued to many better-performing systems.
Fusing the output of a one-shot query issued to many IR systems has
also received attention -- for example, {\citet{vogt2000much}}
empirically shows that there is an implicit upper-bound of systems
that should be fused before diminishing returns on effectiveness are
experienced.
{\citet{vogt1999adaptive}} make a case for fusion in web search based
on parallelizing ``fast but inaccurate IR systems'', and then
combining the lists to obtain results commensurate with a single
high-performance system.
Although fusion of results returned from different commercial web
search-engines can be effective in
practice~{\citep{glover1999architecture}}, the approach is generally
perceived as being inefficient.

Here we introduce a new approach to efficient online re-ranking which
directly leverages data fusion.
The key idea is to amortize the cost of fusion, and use a
pre-processing phase that computes query cluster centroids.
In addition, to efficiently compute offline clusters, we introduce a
cost-sensitive fusion technique that employs a single heap to
simultaneously evaluate all query variations.
Similar queries can be mined or generated {\cite{bcglm18-desires}},
then aggregated together using a variety of well-known techniques
{\citep{wen2001clustering, metzler2007similarity, scholer2002query}}.
For reproducibility, the experiments we report here employ the
publicly available UQV100 test collection~{\citep{bmst16sigir}}.

\begin{figure}[t]
\includegraphics[width=0.6\columnwidth]{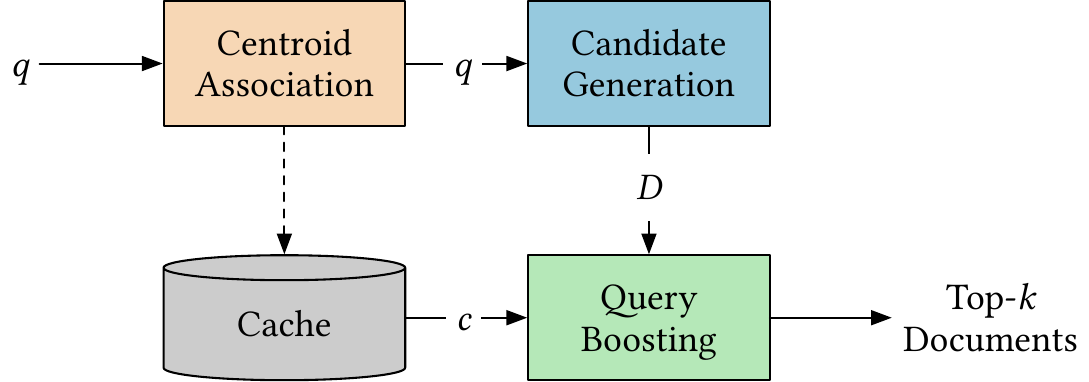}
\mycaption{
The user enters a query $q$, which is both evaluated against the
document collection, and used to search a cache of query clusters.
If a match is found in the cache, cluster centroid $c$ and the
top-$k$ document list $D$ are combined by the query boosting method,
yielding a final top-$k$ result page.
\label{fig-motivation}}
\end{figure}

Figure~\ref{fig-motivation} summarizes the proposed architecture.
Pre-computed {\emph{centroid rankings}} for identified query clusters
are held in a searchable cache.
When a new query arrives, an association process identifies a
matching cluster, and the centroid ranking is retrieved; at the same
time, the query is processed to produce a top-$k$ ranking.
The centroid ranking and the query ranking are then fused, and a
combined top-$k$ ranking computed.
The goal is for that fused ranking to represent the combined
consensus of all of the queries in the pre-processed cluster,
together with the documents specifically identified by the query that
was issued.

Given this proposed combination of offline and online processing, we
present several methods for combining the centroid ranking and the
original query ranking.
These include balanced interleaving between the query ranking and the
centroid ranking, as occurs in online retrieval experiments
{\citep{radlinski2008does}}; carrying out a weighted CombSUM
{\citep{fox1994combination}} between the two rankings; and employing
a re-ranking approach in which the common-to-both documents are
placed at the head of the result page, ordered by their position in
the centroid.

\myparagraph{Contributions}

In particular, in this paper we:
\begin{itemize}
\item Investigate the wall-clock time of single-pass rank fusion and the 
parallel method described by {\citet{vogt1999adaptive}};
\item Propose a novel cost-effective single-pass rank fusion technique;
\item Describe a novel query fusion architecture that efficiently re-ranks 
queries in an online setting; and
\item Validate our results using the ClueWeb12B corpus and
the query variations provided by {\citet{bmst16sigir}}.
\end{itemize}

\noindent
In what follows, Section~\ref{sec-relwork} introduces related work;
Section~\ref{sec-methodology} describes the experimental setup;
Section~\ref{sec-realtime} explores a number of techniques for fusing
multiple query variations {\emph{online}}; Section~\ref{sec-online}
shows how the fusion can be computed {\emph{offline}} and proposes
various rankers that can utilize such pre-computed data;
Section~\ref{sec-discussion} outlines some of the shortcomings of our
investigation and outlines future work; and
Section~\ref{sec-conclusion} concludes the article.

\section{Background}
\label{sec-relwork}

\subsection{Rank Fusion}

It has been known for many years that combining the ranked retrieval
outputs of query variations representing the same information need
can improve retrieval effectiveness.
{\citet{bccc93-sigir}} showed a representative sample of topic
descriptions to ten experienced searchers, generating a pool of five
Boolean queries for each topic.
The outputs of the query variations were combined using an unweighted
sum of retrieval scores.
This aggregation of ranked-retrieval scores was later named CombSUM
by {\citet{fox1994combination}} in their investigation into different
rank fusion techniques.
{\citet{bw99trec}} showed that the same technique could be applied to
ranked retrieval.
{\citet{bailey2017retrieval}} studied these phenomena further in
their exploration of fusion and query variation consistency.
Several recent follow-on studies on query variations and rank fusion
have shown that combining multiple individual rankings consistently
boosts effectiveness
{\cite{bc17-adcs,bgm+17-trec,bgm+18-trec,bcglm18-desires}}.

{\citet{kk11sigir}} showed that inter-document similarities can be
used in a fusion framework to reward documents that are similar to
the head of the result list.
Other work has explored the role of fusion in diversification
{\citep{lrr14sigir}}; automatic generation of query perturbations
{\citep{xc13tois}}; the relationship between fusion and clustering
{\citep{kk11sigir}}; and boosting tail query performance using
supervised rank fusion {\citep{hzlm14cikm}}.
No previous studies have explored the resource implications of these
techniques in a large-scale search environment.

The use of supervised rank fusion with query variations has also been
examined.
{\citet{sssc11wsdm}} describe a supervised data fusion method named
LambdaMerge, which optimizes a retrieval effectiveness metric based
on fusion over user query reformulations and a wide range of document
features.
{\citet{lee2015optimization}} extend the LambdaMerge framework from
data fusion to collection fusion, where query-list features in a
collection are averaged and used as query-vertical features.
These approaches, as well as those of {\citet{hzlm14cikm}}, are
closely aligned with our own.
Other fusion techniques are also
possible~{\cite{anava2016probabilistic,oc18-tutorial}}.

The relationship between supervised fusion and learning-to-rank is an
important issue in its own right, but one which is orthogonal to this
work.
Our approach here does not require supervised learning to achieve
competitive results against strong baselines that were selected based
on their well-established ability to efficiently and effectively rank
documents in a web search environment.
We will explore supervised approaches in future work.

{\citet{xc13tois}} show that query perturbations can be generated
using automatic methods with some success.
{\citet{sssc11wsdm}} describe a supervised data fusion method named
LambdaMerge, which optimizes a retrieval effectiveness metric based
on fusion over user query reformulations and a wide range of document
features.
{\citet{lee2015optimization}} extend the LambdaMerge framework from
data fusion to collection fusion, where query-list features in a
collection are averaged to be utilized as query-vertical features.
{\citet{lrr14sigir}} also consider fusion, showing that it helps with
diversification when queries have multiple interpretations.

\subsection{Efficient Index Traversal} 

The most commonly used structure for top-$k$ document retrieval is
the {\emph{inverted index}}.
Each unique term $t$ is represented by a {\emph{postings list}}, a
sequence of document identifier (docid)/term-frequency
($d_{t,i},f_{t,i}$) pairs, one for every document in which term $t$
appears, where $d_{t,i}$ is the docid of the $i$\,th document
containing $t$, and $f_{t,i}$ is the corresponding within-document
term frequency.
Inverted indexes provide efficient and scalable access to the
necessary statistics for document ranking~{\citep{zm06compsurv}}.
When a query is received, the postings lists associated with the
query terms are fetched, and combined to rank and return the top-$k$
documents.

The way in which the postings lists are iterated, known as the
{\emph{index traversal strategy}}, has a large impact on efficiency,
and different postings layouts are amenable to different traversal
strategies.
Here we focus on the popular {\emph{document-ordered}} index layout,
which is most commonly used with {\emph{Document-at-a-Time}} (\daat)
query processing strategies such as {\wand}~{\cite{zchsz03cikm}} and
more recent block-based variants ({\bmw})
{\cite{dt11-sigir,ccv11-icde,dns13-wsdm,mo+17-sigir}}.
These approaches tend to be more efficient than are {\daat}
{\maxscore} and {\emph{term-at-a-time}} (\taat) approaches
{\cite{tf95ipm,stc05-sigir}}, particularly for short queries, the
most common scenario in web search.
However, for long queries or large candidate sets, the case is less
clear-cut {\citep{fj+11-pvldb,cc+17-wsdm,mcbcl18-wsdm}}; moreover,
fusion over query variations often leads to very long queries.
That is, both {\wand} and {\maxscore} have advantages as well as
disadvantages depending on the length of the query, the number of
postings to process, the term selectivity, and a range of other
factors.

\subsection{Batch Query Processing}

Batch processing is a commonly used technique in the database
community~{\cite{ccbs18-tsas}}, but is exploited less often in web
search.
The majority of queries received by a search system are from
interactive users and must be processed efficiently, with query
latency an important contributor to user
satisfaction~\cite{ba+17-tois}.
{\citet{dabs11-wsdm}} exploit the existence of a subclass of queries
that can be processed in batches to minimize the cost of search.
These queries include operations such as cache updates, internal
testing, and index mining.
{\citeauthor{dabs11-wsdm}} also show how batches of queries can be
processed in the context of large-scale search systems, optimizing
I/O and CPU costs.
Their key observation is that if queries can be batched, then
reordering allows common intersections to be cached, reducing net
evaluation cost.
The task of pre-computing an inverted index is also, of course, a
batch-processing operation.
It is the existence of a pre-computed index that allows queries to be
resolved within millisecond response times over massive document
collections, and has allowed web search to become the critically
important tool that it now is.

\subsection{Caching for Large-Scale Search}

In order to consistently meet {\emph{service level agreements}}
(SLAs), search engines must of necessity avoid redundant computation.
Caching is a simple and common approach for increasing query
throughput at the cost of additional space consumption.
Caches can be deployed at many levels of storage, including in-memory
or on-disk~\cite{wl+14-tois,bg+08-tweb}.
In IR there are two major approaches to caching.
{\emph{List caching}} involves storing commonly accessed postings
lists in fast-access memory {\cite{wl+14-tois,cj+10-www}}.
For example, if the postings lists making up the index are stored on
a SSD, a list caching strategy may opt to keep to $n$ most accessed
postings in main-memory.
Alternatively, {\emph{result caching}} involves storing a query along
with the relevant results (or some proxy thereof) that were returned
for the query~\cite{fagni2006boosting, gt09-www}.
In practice, both list and result caching are useful, and are
deployed in tandem~\cite{bg+08-tweb,wl+14-tois}.
Most caches utilize historical data such as static query logs or
sliding windows of recent queries to build models of {\emph{what}} to
cache, and {\emph{when}} to cache it, and can also be personalized on
a per-user basis~\cite{mw12-sigir}.

The architecture proposed in Figure~\ref{fig-motivation} provides
another form of caching to allow subsequent operations to be made
more efficient.
We explain that architecture more fully in Section~\ref{sec-online}.

\section{Methodology}
\label{sec-methodology}

Before providing details of the new techniques in
Section~\ref{sec-realtime} and~\ref{sec-online}, we first describe
the experimental framework that is employed.

\subsection{Hardware and Software}

Our experiments are conducted on an idle Red Hat Enterprise Linux
Server with $256$ GiB of RAM and two Intel Xeon E5-2690 v3 CPUs, each
with $12$ physical cores.
All algorithms were implemented with C++11 and compiled with GCC
6.3.1 using the highest optimization settings.
Where multi-threading was used, up to $48$ threads were spawned using
the C++ STL threading libraries.
All algorithms were implemented as components within the
state-of-the-art {\vbmw} code-base described by
{\citet{mo+17-sigir}}\footnote{\url{https://github.com/rossanoventurini/Variable-BMW}},
and in support of reproducibility, are also made available for others
(see Section~\ref{sec-soft}).

\subsection{Collections and Indexes}

\begin{table}[t]
\mycaption{The ClueWeb12B and UQV100 resources used.
Note that the queries were stopped and Krovetz stemmed, reducing the number
of distinct queries compared to that reported by {\citet{bmst16sigir}}.
}
\label{tbl-collection}
\begin{tabular}{l r}
\toprule
Documents & $52{,}343{,}021$ \\
Topics & $100$\\ 
Total queries & $10{,}835$\\
Unique queries & $4{,}175$\\
Mean unique queries per topic & $41.75$\\
Hold-out queries & $500$\\
\bottomrule
\end{tabular}
\end{table}

We conduct our experiments across the $52$ million document
ClueWeb12B corpus and employ the UQV100 query collection
{\citep{bmst16sigir}} and its $100$ single-faceted topics derived
from the multi-faceted TREC 2013 and 2014 Web Tracks.
The collection contains $10{,}835$ query variations, sourced from
crowd-workers who were presented with a narrative ``backstory'' for
each topic, and asked to formulate a query in response.
For more details in regard to the size and homogeneity of the
clusters we refer the reader to {\citet{bmst16sigir}}.
{\citet{moffat16adcs}} has also explored properties of this
collection; and {\citet{mbst17acmtois}} discuss the wider
implications of query variations.

To support the required experiments, we split the $10{,}835$ UQV100
queries into two sets: a {\emph{training}} set, used to build the
query variation clusters, and a {\emph{testing}}, or
{\emph{hold-out}} set, used to measure the final performance of the
proposed approaches.
The hold-out set was created by selecting five unique query variants
per topic (that is, queries appearing only a single time in the
UQV100 set), yielding a set of $500$ queries across the $100$ topics.
Each hold-out query was drawn randomly from the corresponding topic's
single-instance variations, without replacement.
This hold-out approach differs from the simple hold-out method
described by {\citet{fuhr2018some}}, as all baselines and new
techniques are evaluated against the total universe of topics, with
at least five query impressions per-topic.
The arrangement also avoids the limitations observed in other train-test
splits in a single-query-per-topic evaluation scenario, where results
are biased by the topic-effect of the generated split.
The training set here represents the most commonly seen query
variations for each topic, and hence can be regarded as being
representative of what could be mined from logs in a production
system.
Table~\ref{tbl-collection} summarizes the situation.
Note that the number of distinct queries is less than in the
underlying UQV collection {\citep{bmst16sigir}} because of our use of
a Krovetz stemmer and a stop list.

We used Indri
5.11\footnote{\url{https://www.lemurproject.org/indri.php}} to index
the collection, and then converted the inverted index into the format
expected by the {\vbmw} code-base.
Before building the {\vbmw} index, we reordered the docid space
using the recursive graph
bisection\footnote{\url{https://github.com/mpetri/recursive_graph_bisection}}
approach of {\citet{dk+16-kdd}}, as it has been shown to
substantially improve index compression.
The average block size of our {\vbmw} index is approximately $40$
integers per block, a result of binary searching for the parameter
$\lambda$ as discussed by {\citet{mo+17-sigir}}.
The final index was compressed using the Partitioned Elias-Fano
mechanism~{\cite{go14-sigir}}.

\subsection{Evaluation Metrics}

We use two metrics to evaluate effectiveness: the recall-based NDCG
approach {\cite{jk02acmtois}} at a fixed cutoff depth of $10$; and
the utility-based RBP method {\cite{mz08acmtois}} applied to each
full ranking, with persistence $\phi=0.8$ and hence an expected
viewing depth of five.
These metrics and cutoffs were selected based on the judgment depth
of the UQV100 collection~\cite{lmc16irj}, and result in relatively
low score uncertainty.

Significance is computed using the Bonferroni-corrected paired
$t$-test, and is denoted by $\dag$ for $p<0.05$, and by $\ddag$ for
$p<0.001$.
Significance was always tested with respect to the strongest
available baseline.

\section{Real-Time Fusion of Query Variants}
\label{sec-realtime}

This section considers the cost of computing fused answer rankings at
query time, assuming (through this section) that it is to be done
without the use of the pre-computed centroids that were foreshadowed
in Figure~\ref{fig-motivation}.

\subsection{Efficiently Processing Variations}
\label{sec-make-centroids}

Given a cluster of queries, including the query just entered by the
user, the goal is to process them all, fuse their results, and return
a single SERP (search engine results page), all the while noting that
web-search systems typically impose strict per-query resource budgets
{\cite{sk15-wsdm,db13-cacm,yh+15-sigir}}.

\myparagraph{Parallel Fusion \mbox{(PF)}}

The simplest approach is to spawn a process for each unique query
variation in the cluster, and execute the queries in parallel.
Once all threads have returned their top-$k$ results, a rank-fusion
algorithm is applied, to assemble the final SERP.
This approach is viable provided there are sufficient CPU cores
available, and requires that each thread operate to the same
response-time requirement as the original query.
If latency as a critical component of the SLA, queries that run
longer than a fixed time threshold could be abandoned and not
considered for the fusion process {\cite{yh+15-sigir,sk15-wsdm}},
thereby sacrificing effectiveness to stay within the resource
constraint.
We measured three variants of the parallel fusion approach, denoted
``\opstyle{PF-}$a$'', where $a$ is a query processing strategy, one
of {\vbmw}, {\wand}, and {\maxscore}.

\myparagraph{Single Pass {\daat}}

A drawback of the parallel approach is that many similar queries are
processed concurrently, and hence that some of the corresponding
postings lists are processed many times, without any commonality
being exploited.
An alternative is to perform all scoring operations in a single
{\daat} pass across the inverted index, concurrently building a
top-$k$ heap for each unique query variant.
That is, an empty top-$k$ heap is constructed for each unique query
variant, and the postings lists for all terms are iterated in
parallel, selecting as the {\emph{pivot}} the minimum document ID
across all of the cursors.
At each processing step, all postings lists are advanced to align
with the pivot, with variables tracking the current set of scores of
the pivot document with respect to the terms appearing in each query.
Once all aligned postings have been processed, the document scores
are checked against the corresponding heaps, each of which is updated
if necessary.
Finally, when all postings cursors are exhausted, the set of heaps
contain the top-$k$ results for the set of query variations, and can
be fused to create the required single SERP.
We refer to this approach as ``{\opstyle{SP-Exhaustive}}''.

\myparagraph{Single Pass {\combsum}}

We now propose an efficient single-pass approach for computing the
{\combsum} {\cite{fox1994combination}} rank fusion score for a set of
query variations, allowing improved efficiency through {\emph{dynamic
pruning}}.
Given a set of ranked lists of documents, and a positive numeric
score for each document in each list, the {\combsum} score for a
document $d$ is the sum of the scores of $d$, computed over its
appearances in the ranked lists.
If there are $\ell$ lists, $L_{1}$ to $L_{\ell}$, and the score of
some document $d$ in the $i$\,th of the lists is given by $\sid$
(with $\sid\equiv0$ if $d\not\in L_i$), then
\[
	\combsum(d) = \sum_{i=1}^{\ell}\sid \, .
\]
Now consider each of the component scores $\sid$, and the query $q_i$
that led to it.
If the scoring computation is an {\emph{additive}} one, then
\[
	\sid = \sum_{t\in q_i} F(t,d) \, ,
\]
where $F(t,d)$ is the term-document score contribution associated with
the term $t$ in the document $d$ according to the chosen retrieval model.
Taking these together gives
\begin{equation}
	\combsum(d) = \sum_{i=1}^{\ell} \left(\sum_{t\in q_i} F(t,d)\right) \,.
	\label{eqn-combsum1}
\end{equation}
Now define $n_t\equiv|\{q_i\mid 1\le i\le\ell\wedge t\in q_i\}|$, the
number of input queries containing term $t$; and $Q\equiv\cup_{1\le
i\le\ell}\,q_i$, the union of the queries.
Equation~\ref{eqn-combsum1} can then be rewritten as
\begin{equation}
        \combsum(d) = \sum_{t\in Q} n_t \cdot F(t,d) \,,
	\label{eqn-combsum2}
\end{equation}
making it clear that for additive similarity scoring mechanisms, the
{\combsum} score for a set of query variations can be computed by
forming the union $Q$ of the queries, counting term frequencies $n_t$
across the variations, and then evaluating a single ``super query''
against the index of the collection and taking a linear sum of the
individual contributions $F(t,d)$.
This requires that all of the scores are positive.
Similarity models that are not additive, or that yield negative
scores, may not be used in this way.
Note also that the usual {\combsum} process of scaling the document
scores into the range $[0\dots1]$ would prevent the simplification
shown in Equations~\ref{eqn-combsum1} and~\ref{eqn-combsum2}.
In the following experiments, we use the BM25 similarity model, and
do not normalize the individual ranking scores.

In order to employ Equation~\ref{eqn-combsum2}, and allow dynamic
pruning techniques to be safely applied to the super query, the index
traversal process must be slightly modified, with the upper-bound
scores, $U_t$, also multiplied by $n_t$.
For the block-based {\vbmw} approach, we must also supply the $n_t$
multiplier to each block-max score, $U_{b,t}$, on-the-fly.
Query processing does not differ in any other way.
This approach is generalizable to all safe-to-$k$ dynamic pruning
traversal strategies; and hence we again test three variations,
denoted ``{\opstyle{SP-CS-}}$a$'', with $a$ one of {\vbmw},
{\wand}, and {\maxscore}.

\begin{figure*}[t]
\includegraphics{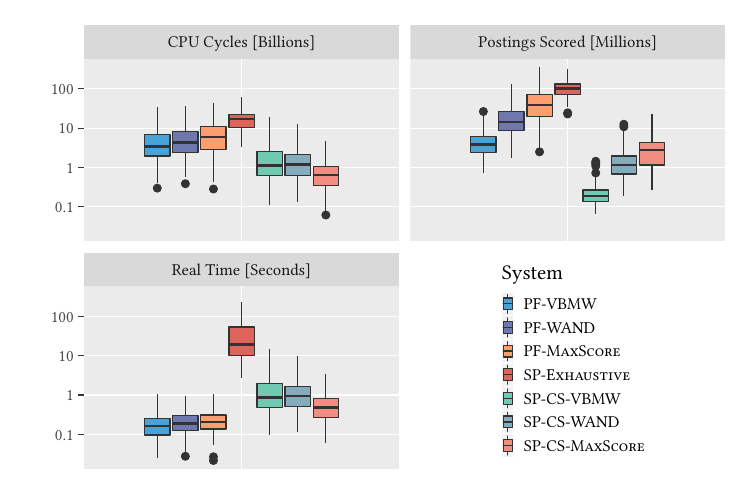}
\mycaption{Efficiency of rank-fusion algorithms, assuming that the
starting point is a set of query variations.
The panes show total cost in CPU cycles; total cost in terms of
postings processed; and query latency.
\label{fig-rf-efficiency}}
\end{figure*}

\subsection{Experiment: Real-Time Fusion}

To test these approaches, we take all query variations for each
topic, and measure the cost of computing a fused top-$100$ result
list.
This involves combining $42$ query variations per topic on average,
each computed to depth $k = 1{,}000$, and then (except in the case of
the {\opstyle{SP-CS}} approaches) fusing the results.
Three indicators are reported: the number of CPU cycles consumed; the
number of postings scored; and response latency.
In the case of the parallel approaches, the first two are summed over
all threads.
Note that the CPU measurements ignore the slight processing overhead
generated by the forking and locking activities inherent in parallel
execution.

Figure~{\ref{fig-rf-efficiency}} shows the results.
The first (top left) pane shows the total CPU time required.
The three {\opstyle{SP-CS}} methods are the most efficient in terms
of processing cost, with the {\opstyle{SP-CS-\maxscore}} approach
slightly better than the other two.
In the second pane (top right), the {\opstyle{SP-CS-\maxscore}}
implementation processes more postings on average than either the
{\opstyle{SP-CS-\wand}} and {\opstyle{SP-CS-\vbmw}} approaches -- the
latter two reduce the number of postings, at the cost of more
non-posting processing.
Finally, the third pane (bottom left) shows elapsed wall-clock time.
The three {\opstyle{PF}} approaches are the fastest, due to their
extensive use of parallelism.
Even though each query's latency is dictated by the slowest-running
variation, execution on average is fast; and the use of suitable
aggregation policies {\cite{yh+15-sigir}} can improve the
{\emph{tail-latency}}~\cite{m17-sigir} of such approaches.
Note, however, that this speed comes at a resource cost, as shown in
the first pane.
If latency and overall workload are both critical concerns,
parallelism should instead be added to the {\opstyle{SP-CS}}
versions, by splitting the collection across the pool of available
processors.

\subsection{How Many Queries Should Be Fused?}
\begin{figure}[t]
\includegraphics{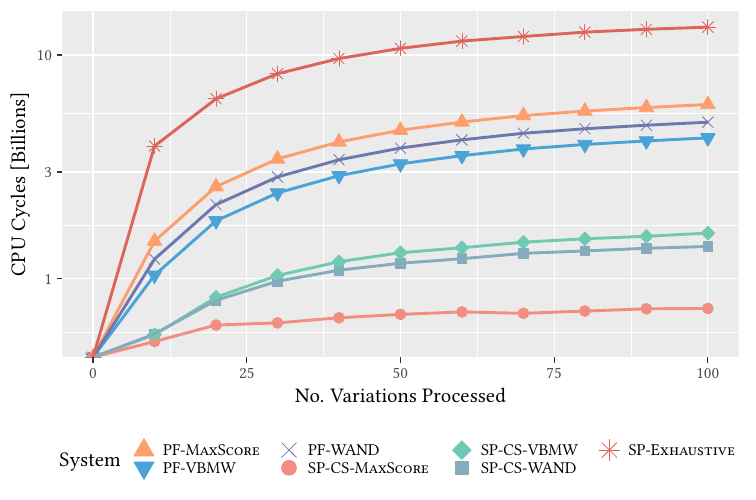}
\mycaption{Average per-topic cost in CPU cycles of approaches for
generating fused rankings, as a function of the number of query
variations being fused.
Sampling is random from each cluster, with replacement,
averaged over ten independent sequences.
All of these methods obtain the same effectiveness, as
listed in Table~\ref{tbl-how-many-effectiveness}.
Note the logarithmic vertical scale.
\label{fig-cost-vs-variants}}
\end{figure}

Another way of reducing processing costs is to fuse fewer queries.
To determine the impact that the number of variants has on processing
costs, random samples (with replacement) were drawn from each UQV100
query cluster.
Those samples were then executed and fused, recording the cost in CPU
cycles and the effectiveness of the final SERP.
The selection process was carried out incrementally, with
one variation drawn and measured, then a second added to it and
measured, and so on; with that entire sequence repeated ten times,
and the values recorded being the averages over those ten runs.
Table~\ref{tbl-how-many-effectiveness} and
Figure~\ref{fig-cost-vs-variants} show the results.

\begin{table}[t]

\mycaption{Fusion effectiveness as the number of query variants is
increased.
Variants are selected from the query clusters at random, with
replacement; measured as averages over a set of ten such sequences.
The values in parentheses are {\metric{RBP}} residuals, recording the
maximum extent of the {\metric{RBP}} score uncertainty.}
\label{tbl-how-many-effectiveness}
\centering
\begin{tabular}{@{}cccc@{}}
\toprule
Num. variants
	& \metric{NDCG$@10$}
		&\metric{RBP $\phi = 0.8$}
\\
\midrule
$1$    & $0.182$ & $0.426~(+0.067)$ \\
$2$    & $0.210$ & $0.469~(+0.081)$ \\
$5$    & $0.236$ & $0.514~(+0.043)$ \\
$10$   & $0.254$ & $0.537~(+0.028)$ \\
$20$   & $0.256$ & $0.540~(+0.021)$ \\
$50$   & $0.261$ & $0.542~(+0.018)$ \\
$100$  & $\mathbf{0.262}$ & $\mathbf{0.550}~(+0.017)$ \\
\bottomrule
\end{tabular}
\end{table}

Table~\ref{tbl-how-many-effectiveness} makes it clear that adding
variants increases effectiveness, but that the gains diminish as more
variants are added.
This outcome is at least partially a consequence of the methodology
used, since ``with replacement'' means that there is an increasing
probability of a previously-selected query being drawn again.
{\citet{bailey2017retrieval}} also observe that, in general, adding
more distinct variants improve the effectiveness of rank fusion,
irrespective of metric or fusion mechanism.

Figure~{\ref{fig-cost-vs-variants}} shows that
{\opstyle{SP-CS-\maxscore}} retains its computational advantage
across the range of variation set sizes, and that all of the
{\opstyle{SP-CS}} approaches are cheaper than the
{\opstyle{SP-Exhaustive}} and {\opstyle{PF}} methods.
The reason for this advantage is that the {\opstyle{SP-CS}}
approaches are largely determined by the number of unique
{\emph{terms}} that are required for processing, whereas the other
methods are more closely tied to the number of unique
{\emph{queries}}.
Since query variations often contain many similar terms, the
{\opstyle{SP-CS}} approaches scale better.

\subsection{Discussion}

Three approaches for efficiently fusing a set of query variations
have been described and measured.
The {\opstyle{SP-Exhaustive}} and {\opstyle{PF}} approaches can be
used with any rank-fusion algorithm, but are more costly than the
{\opstyle{SP-CS}} approaches, which are based on {\combsum} and
additive functions such as BM25.
The {\opstyle{PF}} approaches have the lowest latency, but use nearly
an order of magnitude more resources than the implementation of the
{\opstyle{SP-CS}} methods that was measured here.
As a result, the {\opstyle{SP-CS}} approaches are more scalable with
respect to the number of query variations, and, as already noted, can
be parallelized to reduce their latency.
That is, rank fusion across query variations can be computed such
that latency is small enough for online use, but only if substantial
total computational resources are available.
Better techniques are required if the overall resource cost must also
be managed carefully.

Although this section has focused on fusion using the cost-effective
{\combsum} method described above, there are other fusion techniques
that can also be applied, some of which are easier than others to
transform into single-pass implementations.
{\citet{bc17-adcs}} compare the effectiveness of {\combsum} with six
other unsupervised fusion techniques on the ClueWeb12B corpus using
the UQV100 query set, and demonstrate that {\combsum} is competitive
with the CombMNZ {\cite{fox1994combination}}, RBC
{\cite{bailey2017retrieval}}, and RRF {\cite{cormack2009reciprocal}}
approaches.
The next section shows how even more complex combinations of fusion
over query variations and systems can produce highly effective search
results.

\section{Using Pre-Computed Centroids}
\label{sec-online}

The previous section introduced a reduced-cost query fusion technique
and showed that it is more efficient than computing a separate ranked
list for each query variation.
Even so, if aggressive service level agreements governing response
time and throughput must be complied with, on-the-fly query fusion
may not be a viable approach.
The key issue then becomes: is it possible to compute fused lists in
an offline manner, and use those cached intermediate results to boost
online query performance?
With this question in mind, we now turn to the arrangement that was
sketched in Figure~\ref{fig-motivation}, in which pre-generated query
centroids are stored, and used during online query processing to
adjust the ranking generated for each individual query as it is
executed.

\subsection{Computing Query Centroids} 

Fused centroid lists can be pre-computed using the methods described
in Section~\ref{sec-realtime} at the same time as pseudo-documents
based on the union of the terms in each cluster are constructed and
themselves indexed.
Since this is a pre-computation, total resource requirement is the
appropriate cost measure, and not latency; and with the cost equation
further moderated by the expectation that the pre-processing time can
be amortized over multiple subsequent queries that refer to that
cluster.
Hence, any desired fusion technique can be used, with no restriction
to {\opstyle{BM25}} and/or {\combsum}.

If query centroid data in the form of indexed pseudo-documents and
consensus fused rankings are stored in main memory or on SSD, it can
be searched and retrieved quickly as queries are processed.
In the experiments described shortly, the fused query consensus
rankings (the cluster ranking {\emph{centroids}}) are stored in main
memory, using a doubly-linked circular list that preserves the ranked
order of the fused set; with document identifiers also maintained in
an $\mathcal{O}(1)$ average-time hash table.
With these structures, the UQV100 data required {\kb{49.1}} per topic
on average to store a ranked list containing $k=1000$ documents 
(centroid lists were always computed and stored to a length
of $1{,}000$ documents).
That space requirement can be reduced to {\kb{11.5}} per topic if the
subsequent fusion is restricted to methods that are rank-based, where
knowledge of document identifiers and ranks alone is sufficient
without document scores.

\subsection{Boosting Effectiveness Using Centroids}
\label{sec-online-boosting}

Three different approaches to joining the centroid ranking and a
query ranking -- with the goal of boosting overall effectiveness --
were explored and measured.
While not an exhaustive list of possibilities, these three methods
demonstrate that improvements in risk-sensitivity and retrieval
effectiveness are achievable using relatively simple and inexpensive
techniques.

\myparagraph{Plain Interleaving}

The first approach is inspired by online retrieval experiments in
which results from two different systems are presented as a single
list.
In a balanced interleave, elements are chosen in alternating fashion
from two results, with the first chosen randomly.
Instead, we bias towards the query centroid and always take it first,
assuming that it aggregates more information about the information
need, and is more likely to have high-ranking relevant documents.
We denote this approach as ``{\opstyle{Interleave}}'', and at each
step the next highest ranked document not already included in the
output list is taken from one of the two lists being joined in a
strictly alternating pattern, and added to the fused run.
A possible benefit of the interleaving model is its connection to A/B
testing, where click logs may be helpful in deciding whether to
preference should be given to the user query or the fused result
centroid.

\myparagraph{Linear Combination}

Another option is to adopt the approach proposed by
{\citet{vc99irj}}, and compute a per-document weighted sum of the
min-max re-scaled centroid set and the re-scaled user query answer
set, then use those scores as a descending-order sort key to form the
SERP.
To implement this approach we applied a weight of $\delta$ to the
query centroid and $(1-\delta)$ to the user query, and set $\delta$
to $0.5$ as a starting point.
An exploration of the impact of varying $\delta$ on retrieval
effectiveness and risk-sensitivity is carried out below; when $\delta
> 0.5$, the method approaches the bias exhibited in the
{\opstyle{Interleave}} method.
Unlike the {\opstyle{Interleave}} method, a linear combination does
not guarantee an equal contribution from either the cluster centroid
or the documents retrieved in response to the actual query.
This is because inclusion in the final results list is sensitive to
both the union of the two sets of documents, and also to the scores
of those documents. We denote this approach as ``{\opstyle{LC}}''.

\myparagraph{Reference List Re-Ranking}

\begin{figure}[t]
\includegraphics[height=5cm]{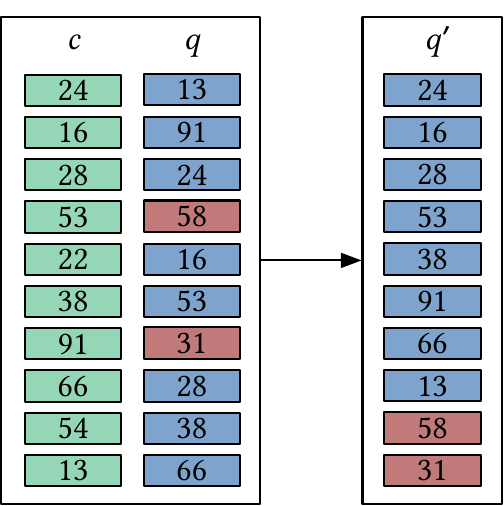}
\mycaption{Reference list re-ranking, where $c$ is the $k$-document
pre-computed centroid ranking, $q$ is the $k$-document run generated
for the user query, and $q'$ is their $k$-document fused result, in
this example with $k=10$ for both of the lists.
Documents in green are from the centroid list, documents in blue are
common to it and the query list, and the documents in red are the
ones from $q$ that were not in $c$.
\label{fig-rerank}}
\end{figure}

This approach intersects the query's ranking and the centroid
ranking, adopting the ordering supplied by the centroid run.
Any remaining documents from the query's run are added after the tail
of the intersection set, in the same order as they appear in the
fused list.
Both the centroid ranking order and the query's ranking order are
respected, but with the documents that were in both listed first.
We denote this approach as ``{\opstyle{Ref-Reorder}}''.
Figure~\ref{fig-rerank} provides an example in which two runs of
length $k=10$ are joined, favoring the documents that appear in both,
and accepting the ordering of the consensus ranking.
Note that in the experiments described shortly the centroid
list was always $1{,}000$ documents long, and only the query's
ranking length was altered during the experimentation.

\subsection{Centroid Construction Approach}

For the following experiments, we construct query centroids by fusing
the training, or ``held-in'', set of queries for each cluster using
.
This allows us to establish how much improvement can be gained
through building clusters using the proposed cost-effective
single-pass query centroid generation scheme introduced in
Section~\ref{sec-make-centroids}.
Note, however, that this is not a necessary part of the arrangement,
and any construction approach can be used to generate centroids, not
just fusing {\opstyle{BM25}} results using {\combsum}.
Other options are discussed in Section~\ref{sec-discussion}.

\subsection{Evaluation}
\label{sec-online-eval}

The baselines used in this section include two proximity-based
algorithms and one learning-to-rank model.

\myparagraph{Term Dependency Models}

We use two term-dependency approaches that represent strong
baselines.
The first is the {\opstyle{L2p}} approach proposed by
{\citet{lmc16ictir}}, a bigram ranking model that linearly combines
the {\opstyle{BM25}} score of a document with sequential bigrams.
Unlike other term dependency models, {\opstyle{L2p}} does not require
global statistics for term dependencies, greatly improving overall
efficiency.
We also use a field-weighted variation of the sequential dependency
model (SDM) {\cite{mc05-sigir}} that operates across the document
body, document title, and inlink text.
For both of these approaches, the ClueWeb09B collection was used in
conjunction with the TREC $2009$--$2012$ Web Track topics to tune the
parameters.

\myparagraph{Learning-to-Rank}

As a learning-to-rank baseline, denoted ``{\opstyle{LtR}}'', the
LightGBM\footnote{\url{https://github.com/Microsoft/LightGBM}}
framework was used to build a LambdaRank model {\cite{brl06-nips}}.
Instead of training on just the hold-out set, we train our model on
both the topics in the hold-out set as well as the most common five
variants from each topic.
This additional exposure leads to a more robust model.
We used ten-fold cross-validation to train and test the LtR model
across a large set of features, including approximately $150$
query-dependent unigram, $150$ query-dependent bigram, $15$ static
document priors, $100$ document-dependent unigram, and $5$
document-dependent bigram features.
The features were generated using the publicly available system
described by {\citet{cgbc17-sigir}}.
We performed an ablation study of the features produced by this
system, and used LightGBM for re-ranking.
An exhaustive description of features and tools to reproduce this
baseline will be released after this paper is accepted.
We refer the interested reader to {\citet{l09-ltr}} and to
{\citet{mso13irj,msoh13acmtois}} for further information on building
a competitive LtR system.

\myparagraph{Risk Sensitivity}

In addition to effectiveness, we consider the risk-sensitivity of all
proposed methods {\citep{dmo14-sigir}}.
Although improved average effectiveness is desirable, this may not be
indicative that the centroid-based boosting algorithm is generalizing
well.
Risk-sensitive retrieval metrics show that a method is improving the
quality of user result pages without making them worse.
Wins, ties, and losses (W/T/L) are shown for each of the approaches
and baselines with respect to the {\opstyle{BM25}} run on the
held-out user query.
Ties are defined to occur if the experimental run score is more than
$10$\% different to the {\opstyle{BM25}} baseline score.
{\citet{dmo14-sigir}} introduced a risk-sensitive retrieval metric
named {\metric{TRisk}} -- a studentized {\metric{URisk}} which
accepts an evaluation metric and a baseline run.
It is computed by providing a run as input, taking the sum of all
improvements over all topics, less the sum of all losses with a
linearly-scaled $\alpha$ value to vary the impact of losses.
{\metric{TRisk}} scores less than $-2$ indicate that the experimental
run exhibits statistically significant harm compared to the baseline, 
and greater than $+2$ indicates an improvement in risk-sensitivity
with statistical confidence.
We use these two measures to evaluate risk-sensitivity, and set the
parameter $\alpha=3$, assessing score decreases to be four times more
damaging than the benefit derived from any numerically identical
score increases.

\subsection{Results} 
\label{sec-online-results}

The methods introduced in the previous section are compared to
baselines in the context of a performance-sensitive query
environment.
Three key usability aspects are monitored -- efficiency,
effectiveness, and risk-sensitivity.

\begin{table}[t]
\mycaption{Mean query boosting time and total query execution time
(milliseconds per query) and percentage overhead due to boosting when
retrieving the top $k$ documents for user queries computed with BM25.
The query centroid rankings were always $1{,}000$ documents long.
\label{tbl-onlinefuse-times}}
\begin{tabular}{llccc}
\toprule
Length
	& Method
		& Boosting
			& Total
				& Overhead \\
\midrule
\multicolumn{1}{r}{$k=10$}
	& \opstyle{Interleave}
		& 1.019
			& 23.479
				& +4.5\%
\\
	& \opstyle{LC}
		& 1.187
			& 23.647
				& +5.3\%
\\
	& \opstyle{Ref-Reorder}
		& \bf 0.092
			& \bf 22.552
				& \bf +0.4\%
\\[1ex]
\multicolumn{1}{r}{$k=100$}
	& \opstyle{Interleave}
		& 1.118
			& 37.126
				& +3.1\%
\\
	& \opstyle{LC}
		& 1.264
			& 37.272
				& +3.5\%
\\
	& \opstyle{Ref-Reorder}
		& \bf 0.147
			& \bf 36.155
				& \bf +0.4\%
\\ [1ex]
\multicolumn{1}{r}{$k=1{,}000$}
	& \opstyle{Interleave}
		& 2.050
			& 68.857
				& +3.1\%
\\
	& \opstyle{LC}
		& 1.999
			& 68.807
				& +3.0\%
\\
	& \opstyle{Ref-Reorder}
		& \bf 0.548
			& \bf 67.355
				& \bf +0.8\%
\\ 
\bottomrule 
\end{tabular}
\end{table}

\myparagraph{Efficiency}

Table~\ref{tbl-onlinefuse-times} provides boosting times for
{\opstyle{BM25}} queries using the three proposed methods, as well as
the end-to-end time and the percentage overhead relative the original
{\opstyle{BM25}} query.
The {\opstyle{Interleave}} and {\opstyle{LC}} approaches show similar
performance,
and when $k=1{,}000$, both techniques take an average of $2$
milliseconds to complete, a $3\%$ overhead on executing the BOW query
using {\opstyle{BM25}}.
The {\opstyle{Ref-Reorder}} approach is approximately four times
faster, taking $0.5$ milliseconds when $k=1{,}000$.
In all cases the end-to-end latency from query submission to final
top-$k$ is dominated by the initial query {\opstyle{BM25}} stage,
which is efficient in practice, and none of the boosting approaches
add more than $6\%$ overhead to the cost of generating the input
query ranking.

\myparagraph{Effectiveness}

Table~\ref{tbl-onlinefuse} lists effectiveness scores for the three
centroid-based fusion methods, and compares them to the four
baseline approaches, using {\metric{NDCG$@10$}} and {\metric{RBP
$\phi=0.8$}} as evaluation metrics, and averaging across the $500$
single-instance held-out queries.
The number of wins, ties, and losses compared to a {\opstyle{BM25}}
evaluation of the same $500$ queries is also reported, together with
{\metric{TRisk}}$_{\alpha=3}$, which penalizes score losses by a
factor of four compared to score gains.
Recall that {\metric{TRisk}} less than $-2$ and greater than $+2$
imply statistical significance.

\begin{table}[t]
\mycaption{Effectiveness of online fusion measured using two
effectiveness metrics, compared to three baselines.
Significance is measured with respect to {\opstyle{LtR}}, the
strongest of those baselines.
The scores listed in the ``Mean'' column are averages across the
$500$ held-out query variations, five for each of the $100$ topics;
the ``W/T/L'' numbers are the respective counts of those $500$
queries for which the method exceeds the {\method{BM25}} baseline by
more than $10$\%, is within $10$\% of the baseline, and is less than
$10$\% of the baseline.
}
\label{tbl-onlinefuse}
\begin{tabular}{l l ccc}
\toprule
Metric
	& Method
		& Mean
			& W/T/L
				& \metric{TRisk$_{\alpha = 3}$}\\
\midrule
\multicolumn{1}{l}{\metric{NDCG@10}}
& User Query (\opstyle{BM25})
	& 0.170$^\ddag$
		& ---
			& ---      
\\
& \opstyle{L2p}
	& 0.169$^\ddag$
		& 160/236/104
			& --6.992    
\\
& \opstyle{SDM+Fields}
	& 0.180$^\dag$
		& 247/76/177
			& --6.801   
\\ 
& \opstyle{LtR}
	& 0.200$^{\hphantom{\dag}}$
		& 281/53/166
			& --4.640    
\\ 
& \opstyle{Interleave}
	& 0.223$^\dag$
		& 305/108/87
			& \hphantom{--}\bf 3.829 
\\  
& \opstyle{LC}
	& 0.231$^\ddag$
		& 322/94/\bf{84}
			& \hphantom{--}3.614     
\\  
& \opstyle{Ref-Reorder}
	& \bf 0.243$^\ddag$
		& \textbf{345}/62/93
			& \hphantom{--}0.812     
\\[1ex]
\multicolumn{1}{l}{\metric{RBP $\phi = 0.8$}}
& User Query (\opstyle{BM25})
	& 0.401~$(+0.088)^\dag$
		& ---
			& ---
\\
& \opstyle{L2p}
	& 0.400~$(+0.089)^\dag$
		& 136/264/100
			& --8.457    
\\
& \opstyle{SDM+Fields}
	& 0.417~$(+0.153)^{\hphantom{\dag}}$
		& 215/117/168
			& --7.557    
\\ 
& \opstyle{LtR}
	& 0.435~$(+0.205)^{\hphantom{\dag}}$
		& 258/71/171
			& --6.790    
\\ 
& \opstyle{Interleave}
	& 0.486~$(+0.037)^\ddag$
		& 287/150/\bf{63}
			& \hphantom{--}\bf 4.178 
\\  
& \opstyle{LC}
	&  0.504~$(+0.033)^\ddag$
		& 290/130/80
			& \hphantom{--}3.762     
\\  
& \opstyle{Ref-Reorder}
	& \bf 0.527~$(+0.053)^\ddag$
		& \textbf{304}/95/101
			& \hphantom{--}1.454 
\\ 
\bottomrule
\end{tabular}
\end{table}

Of the methods tested, the {\opstyle{Ref-Reorder}} mechanism is the
most effective, with an {\metric{NDCG}} score of $0.243$ and
{\metric{RBP}} score of $0.527$ (and in the latter case with a small
residual, indicating that the judgments are a good fit for the runs
that were scored).
This approach also has the lowest risk-sensitivity among all
centroid-based boosting methods, with only $93$ of the queries
performing worse than the user query alone for {\metric{NDCG}} ($101$
for {\metric{RBP}}); note, however, that these results are contingent
on the correct centroid having been identified, an assumption that is
explored below.

The {\metric{TRisk}}$_{\alpha=3}$ values in
Table~\ref{tbl-onlinefuse} provide a somewhat different picture, and
suggest that {\opstyle{Interleave}} boosting is the most desirable if
losses relative to the baseline {\opstyle{BM25}} approach are
penalized more than gains are rewarded, even though when measured by
average per-query scores it is the least useful boosting approach.

\begin{figure}[t]
\includegraphics{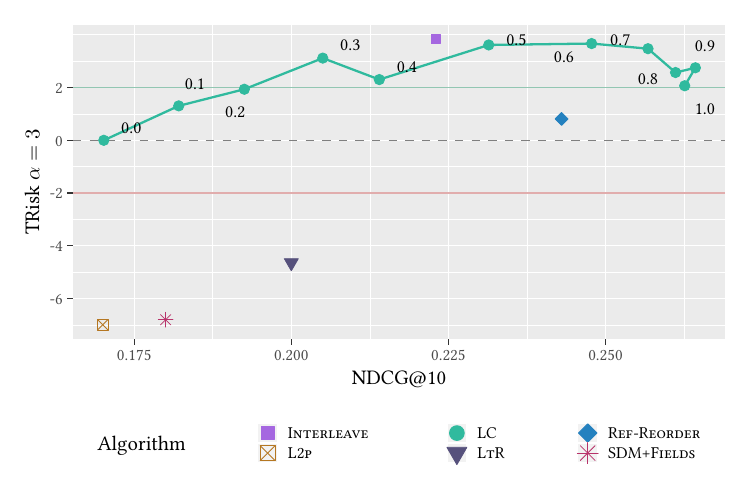}
\mycaption{Varying the weighting of the linear combination (method
{\opstyle{LC}}) between the query centroid and the user's original
query can reduce risk and improve retrieval effectiveness.
The numbers adjacent to the LC line indicate the corresponding values
of~$\delta$; effectiveness tends to be greater when $\delta>0.5$ and
more emphasis is given to the cluster centroid run.
(Recall that {\metric{TRisk}} scores less than $-2$ and greater than
$+2$ indicate statistical significance.)
\label{fig-boost-risk}}
\end{figure}

The {\opstyle{LC}} approach listed in Table~\ref{tbl-onlinefuse}
employed a parameter $\delta$ of $0.5$, placing equal weight on the
two rankings being combined.
Figure~\ref{fig-boost-risk} shows that as $\delta$ is varied the
linear combination of the scores can be adjusted to counterbalance
risk-sensitivity and effectiveness.
For example, shifting to a weighting based $70\%$ on the
query-centroid result set and $30\%$ on the user query set retrieved
by {\opstyle{BM25}} yields an improved risk-reward trade-off compared
to evenly weighting both lists.
Note that the weightings mentioned in Figure~\ref{fig-boost-risk} are
not intended to be prescriptive.
In order to test how generalizable these weights are, query variation
collections formed using a similar methodology to UQV100 would need
to become available to test on other collections.

\begin{figure}[t]
\includegraphics{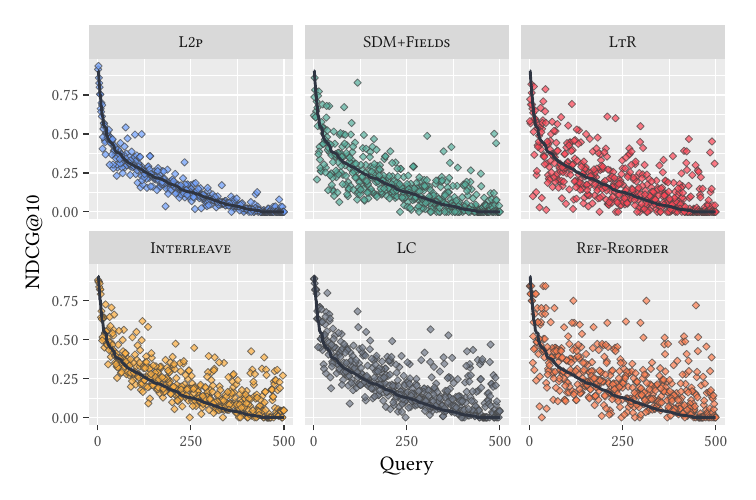}
\mycaption{Difference between {\metric{NDCG$@10$}} effectiveness
score for the baseline {\opstyle{BM25}} query and for six other
evaluation approaches.
The $500$ held-out queries are ordered by their {\opstyle{BM25}}
scores (the black line of dots in each graph) and the corresponding
scores from each of the different evaluation approaches are shown,
one set in each pane.
\label{fig-wtl-ndcg}}
\end{figure}

Figure~{\ref{fig-wtl-ndcg}} provides another perspective on these
issues, and shows the extent to which
{\metric{NDCG}@10} shifts up and down compared to the baseline
{\opstyle{BM25}} ranking on a query-by-query basis.
The top three panels represent effectiveness profiles for the three
other baseline systems, while the lower three panels demonstrate the
profiles of the three boosting techniques described here.
The same decreasing-score ordering of queries is used in each of the
six panes, with the {\opstyle{BM25}} scores shown as black dots and
the corresponding scores for that method shown as colored dots, and
with (in all of the panes) the five other scores for each query shown
as light grey background points.
Of the three baseline systems, {\opstyle{L2p}} follows the
{\opstyle{BM25}} line closely, as it is derived directly from
{\opstyle{BM25}} with a small additional weight attributed to the
sequential dependency scores.
The {\opstyle{SDM+Fields}} approach uses a language model and does
not always agree with {\opstyle{BM25}}, but remains a fair
comparison.
The {\opstyle{LtR}} effectiveness with respect to {\opstyle{BM25}}
exhibits higher variance the other two baselines, presumably since
optimization over many features can induce bigger wins, but can also
bigger losses too.
For the three boosting methods, {\opstyle{Interleave}} and
{\opstyle{LC}} are operating with similar risk-sensitivity, as
evidenced by their similar {\metric{TRisk}} scores in
Table~{\ref{tbl-onlinefuse}}.
{\opstyle{Ref-Reorder}} wins most consistently, but is also vulnerable
to quite notable score degradations compared to
{\opstyle{Interleave}} and {\opstyle{LC}}, and explains why its
{\metric{TRisk}} score is lower.
A similar pattern of performance was observed when the same plots
were generated for {\metric{RBP}}.

\subsection{Online Re-Ranking using Real Queries}
\label{centroid-error}

We now turn our attention to a more fundamental question.
Using pre-computed clusters appears to be a powerful approach if
every query can be neatly categorized against a previously-identified
information need.
But in a real system, a reasonable percentage of incoming queries can
be expected to differ from all known clusters, or worse, to be
matched against the wrong cluster.
How do we handle queries that do not match any cluster, and minimize
the impact on effectiveness of failed query matchings?

\myparagraph{Centroid Identification}

We first ask whether incoming queries can be reliably mapped to an
existing query cluster.
{\citet{wen2002query}} showed that combining query keywords and
cross-reference similarity using document hierarchies from
click-through data can give a precision and recall of approximately
$95$\% in an experiment involving $20{,}000$ query clusters.
A $5$\% error rate is encouraging, given that precision can be traded
against recall, to guard against the potential impact of incorrect
cluster matches.

To replicate this evaluation, we built an index of
{\emph{pseudo-documents}}, each containing the union of all query
terms belonging to one of the UQV100 topics.
To this set of $100$ documents we added $10{,}000$
synthetically-formed query clusters using the anchor text of inlinks
to Wikipedia articles on the ClueWeb09 corpus, adopting the approach
of {\citet{dang2010query}}, who showed that anchor text can be used
as a reliable substitute for user queries.
Each pseudo-document contains the union of the query terms (rather
than allowing duplicates) so as to not bias cluster selection.
Upon receiving a new query, the pseudo-document index is searched
using {\opstyle{BM25}} to find the top-scoring pseudo-document.
A range of policies for forming synthetic centroids from inlink data
were employed to aid with pre-processing, sanitization, and
diversity; code for these steps and a manually curated stop-list are
available in our codebase (see Section~\ref{sec-soft}).
Across the $10{,}000$ clusters, $221{,}989$ queries were used
($41{,}624$ unique); the largest and smallest clusters contained
$442$ queries and $11$ queries respectively.
Finally, to test the effect of increasing the number of
``distractor'' clusters, $0$, $100$, $1{,}000$, $5{,}000$ and
$10{,}000$ were added to the UQV100 clusters, and the success rates
for the five held-out UQV queries per topic were computed, where
success consisted of selecting the correct UQV100 cluster in the
presence of the distractors.
These success rates were $97\%$, $97\%$, $94\%$, $91\%$, and $89\%$
respectively, broadly in line with the results of
{\citet{wen2002query}}.

\myparagraph{Incorrect Centroid Association}

The results in Table~\ref{tbl-onlinefuse} assume that the centroid
identification technique is perfect, which is an unlikely scenario.
To quantify the effect of incorrect matchings, we ran ten trials in
which each query was assigned to the wrong cluster with varying
probabilities, so that the failure profiles of our online combination
approaches can be compared in the presence of more realistic centroid
matching.
We denote the error rate as $\epsilon$ and explore values of
$\epsilon=0.05$ (with 5\% of queries assigned the wrong cluster, the
rate attained by {\citet{wen2002query}}) and a somewhat pessimistic
$\epsilon=0.2$ (with 20\% of queries being misassigned).
The latter is worse than the rate observed in the experiments with
$10{,}000$ distractor centroids reported above.

\begin{table}[t!]
\mycaption{Effectiveness of online fusion approaches, where
$\epsilon$ represents the rate in which incorrect clusters are
selected.
Significance is measured with respect to {\opstyle{LtR}}, the
strongest of the four baselines that were employed.
Return Cluster Centroid (\opstyle{RCC}) is tabulated with a
perfect cluster matching rate of $\epsilon=0.0$ as an initial
reference point.
The scores listed in the ``Mean'' column are averages across the
$500$ held-out query variations, five for each of the $100$ topics;
the ``W/T/L'' numbers are the respective counts of those $500$
queries for which the method exceeds the {\method{BM25}} baseline by
more than $10$\%, is within $10$\% of the baseline, and is less than
$10$\% of the baseline.
\label{tbl-onlinefuse-error}}
\begin{tabular}{@{}llccc@{}}
\toprule
Error Rate
	& Method
		& Mean
			& W/T/L
				& \metric{TRisk$_{\alpha = 3}$}
\\
\midrule
&&{\metric{NDCG@10}}&&
\\
\cmidrule{3-3}
\multirow{1}{*}{$\epsilon=0.0$}
	& \opstyle{RCC}
		& \bf 0.263$^\ddag$
            & {\bf 375}/40/{\bf 85}
                & \bf \hphantom{--}3.075
\\[1ex]
\multirow{1}{*}{$\epsilon=0.05$}
	& \opstyle{RCC}
		& \bf 0.249$^\ddag$
			& 334/63/103
				& \hphantom{--}0.468
\\
	& \opstyle{Interleave}
		& 0.217$^\dag$
			& 298/108/94
				& \hphantom{--}2.221
\\

	& \opstyle{LC}
		& 0.225$^\dag$
			& 309/103/{\bf{88}}
				& \bf \hphantom{--}2.365
\\
	& \opstyle{Ref-Reorder}
		& 0.240$^\ddag$
			& {\bf{343}}/66/91
				& \hphantom{--}0.596
\\[1ex]
\multirow{1}{*}{$\epsilon=0.2$}
	& \opstyle{RCC}
		& 0.209$^{\hphantom{\dag}}$
			& 278/63/159
				& --4.339
\\

	& \opstyle{Interleave}
		& 0.199$^{\hphantom{\dag}}$
			& 277/94/129
				& --2.382
\\
	& \opstyle{LC}
		& 0.203$^{\hphantom{\dag}}$
			& 275/95/130
				& --1.984
\\
	& \opstyle{Ref-Reorder}
		& \bf 0.212$^{\hphantom{\dag}}$
			& {\bf{339}}/75/{\bf{86}}
				& \bf \hphantom{--}0.679
\\[1ex]
\midrule
&&{\metric{RBP $\phi = 0.8$}}&&
\\
\cmidrule{3-3}
\multirow{1}{*}{$\epsilon=0.0$}
	& \opstyle{RCC}
		& \bf 0.553~$(+0.011)^\ddag$
            & {\bf 440}/20/{\bf 40}
                & \bf 11.231
\\[1ex]
\multirow{1}{*}{$\epsilon=0.05$}
	& \opstyle{RCC}
		& \bf 0.523~$(+0.063)^\ddag$
			& 289/101/110
				& --0.356
\\
	& \opstyle{Interleave}
		& 0.472~$(+0.061)^\ddag$
			& 269/163/{\bf{68}}
				& \bf \hphantom{--}1.991
\\
	& \opstyle{LC}
		& 0.489~$(+0.059)^\ddag$
			& 271/141/88
				& \hphantom{--}1.828
\\
	& \opstyle{Ref-Reorder}
		& 0.519~$(+0.056)^\ddag$
			& {\bf{301}}/100/99
				& \hphantom{--}1.384
\\[1ex]
\multirow{1}{*}{$\epsilon=0.2$}
	& \opstyle{RCC}
		& 0.441~$(+0.215)^{\hphantom{\dag}}$
			& 235/78/187
				& --7.302
\\
	& \opstyle{Interleave}
		& 0.435~$(+0.128)^{\hphantom{\dag}}$
			& 222/170/108
				& --4.448
\\
	& \opstyle{LC}
		& 0.444~$(+0.137)^{\hphantom{\dag}}$
			& 218/144/138
				& --4.313
\\
	& \opstyle{Ref-Reorder}
		& \bf 0.498~$(+0.061)^\ddag$
			& {\bf{290}}/119/{\bf{91}}
				& \bf \hphantom{--}1.072
\\
\bottomrule
\end{tabular}
\end{table}

Table~\ref{tbl-onlinefuse-error} shows the {\metric{NDCG}} and
{\metric{RBP}} effectiveness attained in this non-perfect evaluation
scenario, as well as their respective risk-reward trade-offs compared
to a {\opstyle{BM25}} baseline, as quantified by {\metric{TRisk}}.
As a further reference point, Table~\ref{tbl-onlinefuse-error} also
includes the effectiveness score of returning the matched centroid
run {\emph{without}} processing the input query from the user,
denoted as {\opstyle{RCC}} (Return Cluster Centroid).
We also include {\opstyle{RCC}} for the perfect matching scenario
$\epsilon=0.0$, to exhibit how intolerant to failure this strategy is
when faced with the possibility of erroneous cluster matching.
In an ideal world where all possible clusters have been pre-computed
and can be matched perfectly, {\opstyle{RCC}} represents the best
score that could be achieved using the approach shown in
Figure~\ref{fig-motivation}.
The performance of all of the approaches degrades as $\epsilon$
increases, and they are vulnerable to inaccuracies if the clustering
error might be high.
The key point is that if a search engine receives an exact match
well-known query, it can simply return a cached result, which is what
is commonly done in practice~\cite{fagni2006boosting,cj+10-www,
bg+08-tweb, wl+14-tois, gt09-www, mw12-sigir}; what we are exploring
here is the consequence of also allowing approximated matches to be
exploited.

Overall, {\opstyle{Ref-Reorder}} is the most robust technique, as was
also shown in Table~\ref{tbl-onlinefuse}, and remains significantly
better than {\opstyle{LtR}} in terms of effectiveness, even when
$5$\% of queries are assigned to the wrong cluster.
In the two lower halves of the table, with the error rate increased
to $\epsilon=0.2$,
{\opstyle{Ref-Reorder}} remains statistically significantly better
than the {\opstyle{LtR}} baseline. 
In contrast to the results of Table~\ref{tbl-onlinefuse}, this method
now has the most wins and least number of losses of any of the online
fusion approaches considered on both evaluation metrics.
Observe also that the higher error rate leads to increased
{\metric{RBP}} residuals, but with the {\opstyle{Ref-Reorder}}
average residual increasing by least.
Even with unrealistically large error rates such as $\epsilon=0.5$
(not shown in Table~\ref{tbl-onlinefuse-error}), the {\metric{NDCG}}
score for {\opstyle{Ref-Reorder}} is $0.205$ (and for {\metric{RBP}},
is $0.462$~$(+0.075)$).
That is, even with a very high error rate in terms of cluster
identification, retrieval effectiveness remains close to the original
{\metric{BM25}} run prior to the application of boosting
(Table~{\ref{tbl-onlinefuse}}).

\section{Discussion}
\label{sec-discussion}

We have presented three novel query effectiveness boosting
techniques, built around the idea (Figure~\ref{fig-motivation}) of
pre-computed query cluster centroids.
We now consider some of the issues associated with this proposal, and
a range of avenues for possible extension.

\myparagraph{Improving Real-Time Rank Fusion}

Although the real-time rank fusion approaches outlined in
Section~\ref{sec-realtime} can be implemented in a low-latency
manner, they nevertheless require substantial amounts of computation.
For this reason, we opted to build query centroids offline using a
cost-effective rank-fusion approach based on {\combsum}
(Section~\ref{sec-online}), and deploy online boosting approaches
that make use of those centroids.
It would be interesting to further develop these approaches to make
them more scalable and less resource intensive.
One possibility would be to employ large-scale distributed
architectures, with parallel fusion conducted across multiple index
server nodes and multiple (perhaps even all) clusters at a time.
Another interesting avenue for future work is to develop ways in
which non-additive fusion techniques such as
RBC~\cite{bailey2017retrieval} and RRF~\cite{cormack2009reciprocal}
can be incorporated into the same framework.
As a motivation, recent work has shown that cost-effective supervised
rank fusion techniques can outperform state-of-the-art
learning-to-rank models~\cite{mm18-cikm}.

\myparagraph{Simulating Query Intent and Data Privacy}

To demonstrate our results in a laboratory setting, we used the
UQV100 test collection, in which the clustering is a direct
consequence of the data collection process; then, in order to measure
the impact of incorrect cluster identification, we also carried out a
failure analysis (Section~{\ref{centroid-error}}).
Even so, future exploration is required in order to verify the
reliability and robustness of the approaches we propose, including
whether it is helpful to form centroids with respect to the current
step in a user information foraging activity; and how effective the
new techniques are outside of test-collection settings.

To validate our observation against a full-scale search engine, query
logs and click-graphs would be required to form these clusters using
automatic methods.
That data is not currently publicly available, and perhaps never will
be after the concerns raised in connection with Cambridge
Analytica~\cite{cambridgeanalytica} and the earlier AOL log
release~\cite{aolsearch}.
Without such data, academics are constrained to exploring performance
improvements using data that is publicly available, as we have done
here.
The upside of using public resources such as the UQV100 queries -- no
matter how limited they may be in scope -- is that the experiments
are reproducible, and do not rely on access to private information
that may have been gathered from users without adequate consent being
given.
Fortunately, previous research from industry research labs has shown
that query clustering by intent is not only possible, but that it is
being used in several different contexts, which we discuss now.

\myparagraph{Forming Query Centroids Automatically}

A range of authors have carried out experiments in connection with
grouping queries by intent (that is, by information need).
Relevant applications include query
rewriting~\cite{bmc12-wsdm,ht+16-cikm, bcglm18-desires}, and forming
query clusters using query logs and click-graphs~\cite{cs07-sigir,
kjs-16-google, sk+18-cikm}.
A team of Microsoft researchers showed that web document click-graphs
can be used to generate ``virtual'' queries by associating documents
that are semantically close on the click-graph~\cite{xz+04-cikm}.
This technique was used to create gating features for query
variations in LambdaMerge~{\cite{sssc11wsdm}}.
{\citet{cs07-sigir}} further demonstrated that random walks on a
click graph can form effective query clusters in the domain of image
search, noting that this is perhaps not unsurprising, given the prior
work of {\citet{xz+04-cikm}}.

\begin{figure}[t]
\includegraphics[width=0.7\columnwidth]{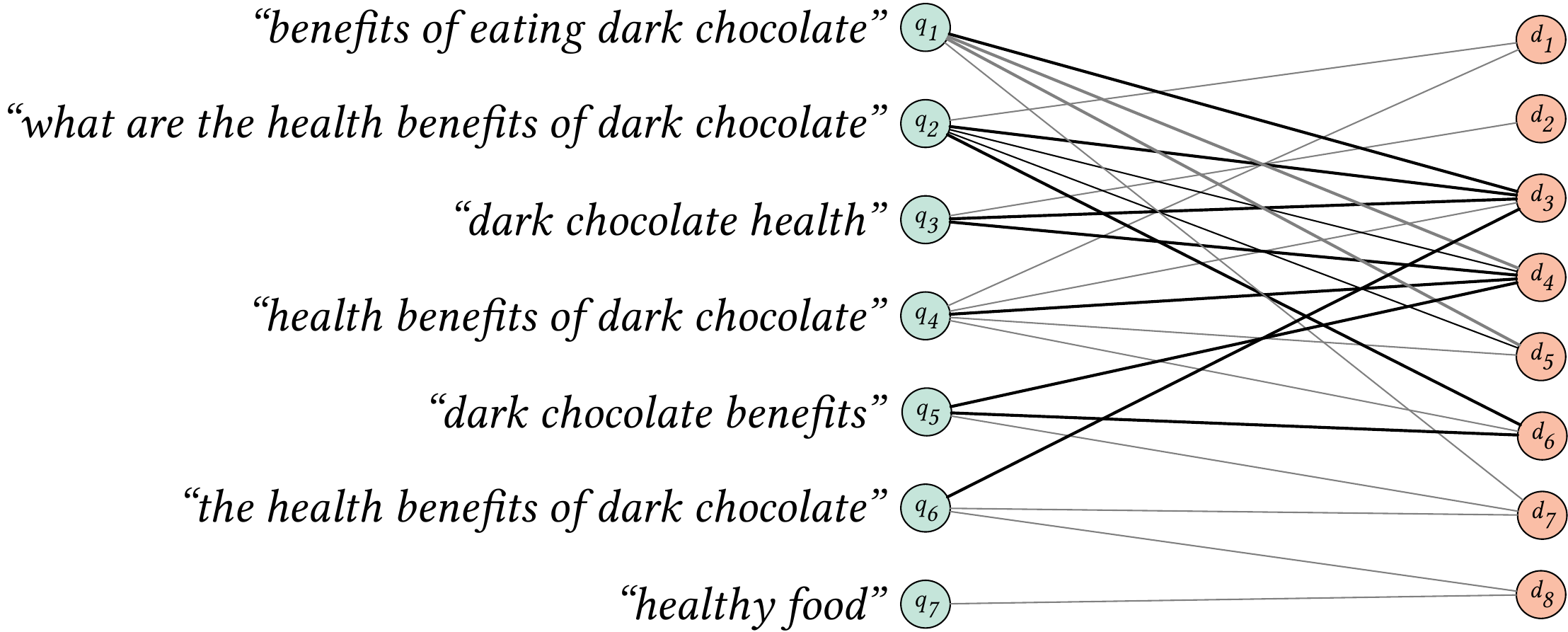}
\mycaption{A bipartite click-graph, showing the associations of document clicks
from queries.
The thickness of each line represents the frequency of clicks for that
query and document pair.
\label{fig-bipartite}}
\end{figure}

Similarly, a team of Google researchers {\citep{kjs-16-google}} write
that:
\begin{quote}\it
Weighted bigraph clustering capitalizes on organic
search results to construct a bipartite graph with a set of queries
and a set of URLs as nodes.
Edge weights of the graph are computed with the impression and click
data of (query, URL) pairs from a Bayesian perspective and are used
to induce query (URL) pairwise similarities.
Due to information embedded in Google search results, this method is
superb in grouping semantically close queries together.
\end{quote}
Figure~{\ref{fig-bipartite}} shows an illustrative click-graph which
can be combined with random walks to induce query variations directly
from large query logs; using such a structure, an at-scale exploration
of the ideas we have introduced here would be a useful further step
in terms of validating the new approach.

\myparagraph{Query Centroid Trade-Offs}

\begin{table}[t!]
\mycaption{Further effectiveness gains attained using a {\emph{double
fusion}} {\cite{bmst16sigir}} query centroid construction strategy.
The baseline here is taken to be the corresponding boosted mechanism
from Table~\ref{tbl-onlinefuse} using (only) the
{\opstyle{BM25}}-based runs across the training queries;
statistical significance and {\opstyle{TRisk}} in this table
are measured relative to that starting point.
The scores listed in the ``Mean'' column are averages across the
$500$ held-out query variations, five for each of the $100$ topics;
the ``W/T/L'' numbers are the respective counts of those $500$
queries for which the method exceeds the baseline more than $10$\%,
is within $10$\% of the baseline, and is less than $10$\% of the
baseline.
\label{tbl-further-effectiveness}}
\begin{tabular}{@{}llccc@{}}
\toprule
Metric
& Method
	& Mean
		& W/T/L
			& \metric{TRisk$_{\alpha = 3}$}
\\
\midrule
\multirow{1}{*}{NDCG@10}	
&\opstyle{Interleave}
	& 0.233$^\dag$
		& 166/182/152
			& --7.092
\\
&\opstyle{LC}
	& 0.243$^\ddag$
        & {\bf 212}/148/{\bf 140}
			& --6.163
\\
&\opstyle{Ref-Reorder}
    & {\bf 0.252}$^\dag$
        & 182/157/161
			& --7.638
\\[1ex]
\multirow{1}{*}{RBP $\phi = 0.8$}
&\opstyle{Interleave}
	& 0.497~$(+0.037)^\dag$
        & 152/232/{\bf 116}
			& --6.602
\\
&\opstyle{LC}
	& 0.519~$(+0.032)^\dag$
        & {\bf 177}/204/119
			& --6.519
\\
&\opstyle{Ref-Reorder}
    & {\bf 0.539}~$(+0.049)$
		& 176/181/143
			& --8.075
\\
\bottomrule
\end{tabular}
\end{table}

In the current experiments, query clusters were generated through a
simple {\combsum} fusion of {\opstyle{BM25}} ranked lists constructed
from query variations.
This simple approach allows for cost-effective cluster computation,
as explored in Section~\ref{sec-realtime}; with improved query
effectiveness a realized goal.
Curating more refined (and also more expensive) query clusters might
further improve the effectiveness of the proposed boosting
approaches.

In a preliminary test of that hypothesis, we built another set of
query clusters based on the {\opstyle{RBC}} fusion approach of
{\citet{bmst16sigir}}, who considered rank fusion across both query
variations and across different systems -- a methodology they refer
to as {\emph{double fusion}}.
As input, we ran a number of ranking models from Indri, including
{\opstyle{BM25}}, {\opstyle{SDM+Fields}}, {\opstyle{BM25}} with
{\opstyle{RM3}} query expansion, and {\opstyle{SDM+Fields}} with
{\opstyle{RM3}} query expansion.
We also added several models from
Terrier\footnote{\url{http://terrier.org/}}, including
{\opstyle{PL2}}; {\opstyle{PL2}} with {\opstyle{Bo1}} query
expansion; four instances of {\opstyle{BM25}} with different
parameterizations (two with {\opstyle{Bo1}} query expansion, and two
without); and four instances of {\opstyle{InL2}} with different
parameterizations (two with query expansion, and two without).
There were $14$ runs used in total.
The resultant ranked lists constructed from each of the training
query variations associated with each UQV100 topic was fused together
using {\combsum} to create a final query cluster result, and then
measured using the $500$ held-out queries, using the methodology
described in Section~\ref{sec-methodology} and used in
Section~\ref{sec-realtime}.

Table~\ref{tbl-further-effectiveness} shows the effectiveness of
three boosting approaches using these more expensive clusters.
The baseline used to ascertain risk-reward profile and statistical
significance are the corresponding boosted mechanisms presented in
Table~\ref{tbl-onlinefuse}.
Higher quality query centroids translate to improved effectiveness,
validating our hypothesis.
We plan to explore the interesting problem of additivity and fusion
effectiveness across both systems and queries in future work.
Note, however, the large negative values associated with this change,
indicating harm to the baseline for the {\metric{TRisk}} $\alpha=3$
metric when compared to the fused {\opstyle{BM25}} starting point.
The gains recorded are, at this stage, quite unevenly distributed
across topics, and more work is required in this area.

\myparagraph{Beyond Single-Faceted Information Needs}

The UQV100 test collection is, by design, composed of topics for
$100$ single-facet information needs.
Multi-faceted search would invariably provide additional challenges.
If a suitable threshold in cluster association scores cannot be met
to confidently associate a query with a single centroid, a fusion of
many faceted query centroids may be required to resolve the user's
information need.
An alternative approach could involve building diversified query
centroids for information needs that are typically diverse, allowing
diversification to be implicitly included via the mechanisms we have
introduced here.
Search result diversification~{\cite{aghi09-wsdm}} is an important
problem in its own right, and fusion techniques have already been
shown to be highly effective for this problem~{\cite{lrr14sigir}}.
We plan to also explore this problem further in future work.

\section{Conclusions}
\label{sec-conclusion}

We have explored three different strategies to improve search
effectiveness for real-time query streams utilizing pre-computed
query centroids.
In general, centroid-based re-ranking techniques offer a highly
efficient mechanism for boosting query effectiveness -- requiring on
average only $2$ milliseconds to reorder $1{,}000$ documents, while
improving effectiveness significantly.
We further show that on-the-fly rank fusion is viable, and can be
reasonably efficient using current state-of-the-art dynamic pruning
techniques, but if aggressive SLAs on query performance are enforced,
carrying out fusion at run-time remains costly.

We then demonstrated that query level fusion can instead be used to
combine similar queries {\emph{offline}}, making it a practical
alternative in high-performance search engines.
To validate this idea, we explore how query clusters can be used to
improve the effectiveness of incoming queries using three different
approaches, namely re-ranking, interleaving and a linear combination
of the cluster and the user query.
Experiments using the ClueWeb12B UQV100 collection show that the new
approaches we describe provide competitive efficiency, and, at the
same time, effectiveness improvements over strong baselines in a
performance-focused query processing framework.

\myparagraph{Software}
\label{sec-soft}
In the interests of reproducibility, our codebase is 
available at \url{https://github.com/rmit-ir/centroid-boost}.

\myparagraph{Acknowledgements} 
This work was supported by the Australian Research Council's
{\emph{Discovery Projects}} Scheme (DP170102231) and a grant from the
Mozilla Foundation.

\bibliographystyle{acm-reference-fmt}
\setlength{\bibsep}{1.0pt}
\bibliography{strings-long,local}


\begin{thebibliography}{00}


\ifx \showCODEN    \undefined \def \showCODEN     #1{\unskip}     \fi
\ifx \showDOI      \undefined \def \showDOI       #1{{\tt DOI:}\penalty0{#1}\ }
  \fi
\ifx \showISBNx    \undefined \def \showISBNx     #1{\unskip}     \fi
\ifx \showISBNxiii \undefined \def \showISBNxiii  #1{\unskip}     \fi
\ifx \showISSN     \undefined \def \showISSN      #1{\unskip}     \fi
\ifx \showLCCN     \undefined \def \showLCCN      #1{\unskip}     \fi
\ifx \shownote     \undefined \def \shownote      #1{#1}          \fi
\ifx \showarticletitle \undefined \def \showarticletitle #1{#1}   \fi
\ifx \showURL      \undefined \def \showURL       #1{#1}          \fi
\providecommand\bibfield[2]{#2}
\providecommand\bibinfo[2]{#2}
\providecommand\natexlab[1]{#1}
\providecommand\showeprint[2][]{arXiv:#2}

\bibitem[\protect\citeauthoryear{Agrawal, Gollapudi, Halverson, and
  Ieong}{Agrawal et~al\mbox{.}}{2009}]%
        {aghi09-wsdm}
\bibfield{author}{\bibinfo{person}{R. Agrawal}, \bibinfo{person}{S. Gollapudi},
  \bibinfo{person}{A. Halverson}, {and} \bibinfo{person}{S. Ieong}.}
  \bibinfo{year}{2009}\natexlab{}.
\newblock \showarticletitle{Diversifying search results}. In
  \bibinfo{booktitle}{{\em Proc. Conf. on Web Search and Data Mining (WSDM)}}.
  \bibinfo{pages}{5--14}.
\newblock


\bibitem[\protect\citeauthoryear{Anava, Shtok, Kurland, and Rabinovich}{Anava
  et~al\mbox{.}}{2016}]%
        {anava2016probabilistic}
\bibfield{author}{\bibinfo{person}{Y. Anava}, \bibinfo{person}{A. Shtok},
  \bibinfo{person}{O. Kurland}, {and} \bibinfo{person}{E. Rabinovich}.}
  \bibinfo{year}{2016}\natexlab{}.
\newblock \showarticletitle{A probabilistic fusion framework}. In
  \bibinfo{booktitle}{{\em Proc. International Conf. on Theory of Information
  Retrieval (ICTIR)}}. \bibinfo{pages}{1463--1472}.
\newblock


\bibitem[\protect\citeauthoryear{Baeza-Yates, Gionis, Junqueira, Murdock,
  Plachouras, and Silvestri}{Baeza-Yates et~al\mbox{.}}{2008}]%
        {bg+08-tweb}
\bibfield{author}{\bibinfo{person}{R. Baeza-Yates}, \bibinfo{person}{A.
  Gionis}, \bibinfo{person}{F. Junqueira}, \bibinfo{person}{V. Murdock},
  \bibinfo{person}{V. Plachouras}, {and} \bibinfo{person}{F. Silvestri}.}
  \bibinfo{year}{2008}\natexlab{}.
\newblock \showarticletitle{Design trade-offs for search engine caching}.
\newblock \bibinfo{journal}{{\em ACM Trans. on the Web\/}} \bibinfo{volume}{2},
  \bibinfo{number}{4} (\bibinfo{year}{2008}), \bibinfo{pages}{1--28}.
\newblock


\bibitem[\protect\citeauthoryear{Bai, Arapakis, Cambazoglu, and Freire}{Bai
  et~al\mbox{.}}{2017}]%
        {ba+17-tois}
\bibfield{author}{\bibinfo{person}{X. Bai}, \bibinfo{person}{I. Arapakis},
  \bibinfo{person}{B.~B. Cambazoglu}, {and} \bibinfo{person}{A. Freire}.}
  \bibinfo{year}{2017}\natexlab{}.
\newblock \showarticletitle{Understanding and leveraging the impact of response
  latency on user behaviour in web search}.
\newblock \bibinfo{journal}{{\em ACM Trans. on Information Systems\/}}
  \bibinfo{volume}{36}, \bibinfo{number}{2} (\bibinfo{year}{2017}),
  \bibinfo{pages}{1--42}.
\newblock


\bibitem[\protect\citeauthoryear{Bailey, Moffat, Scholer, and Thomas}{Bailey
  et~al\mbox{.}}{2016}]%
        {bmst16sigir}
\bibfield{author}{\bibinfo{person}{P. Bailey}, \bibinfo{person}{A. Moffat},
  \bibinfo{person}{F. Scholer}, {and} \bibinfo{person}{P. Thomas}.}
  \bibinfo{year}{2016}\natexlab{}.
\newblock \showarticletitle{UQV100: A test collection with query variability}.
  In \bibinfo{booktitle}{{\em Proc. ACM Conf. on Research and Development in
  Information Retrieval (SIGIR)}}. \bibinfo{pages}{725--728}.
\newblock


\bibitem[\protect\citeauthoryear{Bailey, Moffat, Scholer, and Thomas}{Bailey
  et~al\mbox{.}}{2017}]%
        {bailey2017retrieval}
\bibfield{author}{\bibinfo{person}{P. Bailey}, \bibinfo{person}{A. Moffat},
  \bibinfo{person}{F. Scholer}, {and} \bibinfo{person}{P. Thomas}.}
  \bibinfo{year}{2017}\natexlab{}.
\newblock \showarticletitle{Retrieval consistency in the presence of query
  variations}. In \bibinfo{booktitle}{{\em Proc. ACM Conf. on Research and
  Development in Information Retrieval (SIGIR)}}. \bibinfo{pages}{395--404}.
\newblock


\bibitem[\protect\citeauthoryear{Barbaro and Zeller}{Barbaro and
  Zeller}{2006}]%
        {aolsearch}
\bibfield{author}{\bibinfo{person}{M. Barbaro} {and} \bibinfo{person}{T.
  Zeller}.} \bibinfo{year}{2006}\natexlab{}.
\newblock \bibinfo{title}{A face is exposed for AOL searcher No. 4417749}.
\newblock
  \bibinfo{howpublished}{\url{https://nytimes.com/2006/08/09/technology/09aol.html}}.
    (\bibinfo{date}{Aug.} \bibinfo{year}{2006}).
\newblock
\newblock
\shownote{Accessed: 2018-11-08.}


\bibitem[\protect\citeauthoryear{Belkin, Cool, Croft, and Callan}{Belkin
  et~al\mbox{.}}{1993}]%
        {bccc93-sigir}
\bibfield{author}{\bibinfo{person}{N.~J. Belkin}, \bibinfo{person}{C. Cool},
  \bibinfo{person}{W.~B. Croft}, {and} \bibinfo{person}{J.~P. Callan}.}
  \bibinfo{year}{1993}\natexlab{}.
\newblock \showarticletitle{The effect of multiple query variations on
  information retrieval system performance}. In \bibinfo{booktitle}{{\em Proc.
  ACM Conf. on Research and Development in Information Retrieval (SIGIR)}}.
  \bibinfo{pages}{339--346}.
\newblock


\bibitem[\protect\citeauthoryear{Bendersky, Metzler, and Croft}{Bendersky
  et~al\mbox{.}}{2012}]%
        {bmc12-wsdm}
\bibfield{author}{\bibinfo{person}{M. Bendersky}, \bibinfo{person}{D. Metzler},
  {and} \bibinfo{person}{W.~B. Croft}.} \bibinfo{year}{2012}\natexlab{}.
\newblock \showarticletitle{Effective query formulation with multiple
  information sources}. In \bibinfo{booktitle}{{\em Proc. Conf. on Web Search
  and Data Mining (WSDM)}}. \bibinfo{pages}{443--452}.
\newblock


\bibitem[\protect\citeauthoryear{Benham and Culpepper}{Benham and
  Culpepper}{2017}]%
        {bc17-adcs}
\bibfield{author}{\bibinfo{person}{R. Benham} {and} \bibinfo{person}{J.~S.
  Culpepper}.} \bibinfo{year}{2017}\natexlab{}.
\newblock \showarticletitle{Risk-reward trade-offs in rank fusion}. In
  \bibinfo{booktitle}{{\em Proc. Australasian Document Computing Symp.
  (ADCS)}}. \bibinfo{pages}{1:1--1:8}.
\newblock


\bibitem[\protect\citeauthoryear{Benham, Culpepper, Gallagher, Lu, and
  Mackenzie}{Benham et~al\mbox{.}}{2018}]%
        {bcglm18-desires}
\bibfield{author}{\bibinfo{person}{R. Benham}, \bibinfo{person}{J.~S.
  Culpepper}, \bibinfo{person}{L. Gallagher}, \bibinfo{person}{X. Lu}, {and}
  \bibinfo{person}{J. Mackenzie}.} \bibinfo{year}{2018}\natexlab{}.
\newblock \showarticletitle{Towards efficient and effective query variant
  generation}. In \bibinfo{booktitle}{{\em Proc. Conf. on Design of
  Experimental Search \& Information Retrieval Systems (DESIRES)}}.
  \bibinfo{pages}{62--67}.
\newblock


\bibitem[\protect\citeauthoryear{Benham, Gallagher, Mackenzie, Damessie, Chen,
  Scholer, Moffat, and Culpepper}{Benham et~al\mbox{.}}{2017}]%
        {bgm+17-trec}
\bibfield{author}{\bibinfo{person}{R. Benham}, \bibinfo{person}{L. Gallagher},
  \bibinfo{person}{J. Mackenzie}, \bibinfo{person}{T.~T. Damessie},
  \bibinfo{person}{R-C. Chen}, \bibinfo{person}{F. Scholer},
  \bibinfo{person}{A. Moffat}, {and} \bibinfo{person}{J.~S. Culpepper}.}
  \bibinfo{year}{2017}\natexlab{}.
\newblock \showarticletitle{RMIT at the 2017 TREC CORE track}. In
  \bibinfo{booktitle}{{\em Proc. Text Retrieval Conf. (TREC)}}.
\newblock


\bibitem[\protect\citeauthoryear{Benham, Gallagher, Mackenzie, Liu, Lu,
  Scholer, Moffat, and Culpepper}{Benham et~al\mbox{.}}{2018}]%
        {bgm+18-trec}
\bibfield{author}{\bibinfo{person}{R. Benham}, \bibinfo{person}{L. Gallagher},
  \bibinfo{person}{J. Mackenzie}, \bibinfo{person}{B. Liu}, \bibinfo{person}{X.
  Lu}, \bibinfo{person}{F. Scholer}, \bibinfo{person}{A. Moffat}, {and}
  \bibinfo{person}{J.~S. Culpepper}.} \bibinfo{year}{2018}\natexlab{}.
\newblock \showarticletitle{RMIT at the 2018 TREC CORE track}. In
  \bibinfo{booktitle}{{\em Proc. Text Retrieval Conf. (TREC)}}.
\newblock


\bibitem[\protect\citeauthoryear{Broder, Carmel, Herscovici, Soffer, and
  Zien}{Broder et~al\mbox{.}}{2003}]%
        {zchsz03cikm}
\bibfield{author}{\bibinfo{person}{A.~Z. Broder}, \bibinfo{person}{D. Carmel},
  \bibinfo{person}{M. Herscovici}, \bibinfo{person}{A. Soffer}, {and}
  \bibinfo{person}{J.~Y. Zien}.} \bibinfo{year}{2003}\natexlab{}.
\newblock \showarticletitle{Efficient query evaluation using a two-level
  retrieval process}. In \bibinfo{booktitle}{{\em Proc. ACM International Conf.
  on Information and Knowledge Management (CIKM)}}. \bibinfo{pages}{426--434}.
\newblock


\bibitem[\protect\citeauthoryear{Buckley and Walz}{Buckley and Walz}{1999}]%
        {bw99trec}
\bibfield{author}{\bibinfo{person}{C. Buckley} {and} \bibinfo{person}{J.
  Walz}.} \bibinfo{year}{1999}\natexlab{}.
\newblock \showarticletitle{The TREC-8 query track}. In
  \bibinfo{booktitle}{{\em Proc. Text Retrieval Conf. (TREC)}}.
\newblock


\bibitem[\protect\citeauthoryear{Burges, Ragno, and Le}{Burges
  et~al\mbox{.}}{2006}]%
        {brl06-nips}
\bibfield{author}{\bibinfo{person}{C. Burges}, \bibinfo{person}{R. Ragno},
  {and} \bibinfo{person}{Q.~V. Le}.} \bibinfo{year}{2006}\natexlab{}.
\newblock \showarticletitle{Learning to Rank with nonsmooth cost functions}. In
  \bibinfo{booktitle}{{\em Proc. Conf. on Neural Information Processing Systems
  (NIPS)}}. \bibinfo{pages}{193--200}.
\newblock


\bibitem[\protect\citeauthoryear{Cadwalladr and Graham-{H}arrison}{Cadwalladr
  and Graham-{H}arrison}{2018}]%
        {cambridgeanalytica}
\bibfield{author}{\bibinfo{person}{C. Cadwalladr} {and} \bibinfo{person}{E.
  Graham-{H}arrison}.} \bibinfo{year}{2018}\natexlab{}.
\newblock \bibinfo{title}{Revealed: 50 million Facebook profiles harvested for
  Cambridge Analytica in major data breach}.
\newblock
  \bibinfo{howpublished}{\url{https://theguardian.com/news/2018/mar/17/cambridge-analytica-facebook-influence-us-election}}.
    (\bibinfo{date}{March} \bibinfo{year}{2018}).
\newblock
\newblock
\shownote{Accessed: 2018-11-08.}


\bibitem[\protect\citeauthoryear{Cambazoglu, Junqueira, Plachouras,
  Banachowski, Cui, Lim, and Bridge}{Cambazoglu et~al\mbox{.}}{2010}]%
        {cj+10-www}
\bibfield{author}{\bibinfo{person}{B.~B. Cambazoglu}, \bibinfo{person}{F.~P.
  Junqueira}, \bibinfo{person}{V. Plachouras}, \bibinfo{person}{S.
  Banachowski}, \bibinfo{person}{B. Cui}, \bibinfo{person}{S. Lim}, {and}
  \bibinfo{person}{B. Bridge}.} \bibinfo{year}{2010}\natexlab{}.
\newblock \showarticletitle{A refreshing perspective of search engine caching}.
  In \bibinfo{booktitle}{{\em Proc. Conf. on the World Wide Web (WWW)}}.
  \bibinfo{pages}{181--190}.
\newblock


\bibitem[\protect\citeauthoryear{Chakrabarti, Chaudhuri, and Ganti}{Chakrabarti
  et~al\mbox{.}}{2011}]%
        {ccv11-icde}
\bibfield{author}{\bibinfo{person}{K. Chakrabarti}, \bibinfo{person}{S.
  Chaudhuri}, {and} \bibinfo{person}{V. Ganti}.}
  \bibinfo{year}{2011}\natexlab{}.
\newblock \showarticletitle{Interval-based pruning for top-$k$ processing over
  compressed lists}. In \bibinfo{booktitle}{{\em Proc. International Conf. on
  Data Engineering (ICDE)}}. \bibinfo{pages}{709--720}.
\newblock


\bibitem[\protect\citeauthoryear{Chen, Gallagher, Blanco, and Culpepper}{Chen
  et~al\mbox{.}}{2017}]%
        {cgbc17-sigir}
\bibfield{author}{\bibinfo{person}{R-C. Chen}, \bibinfo{person}{L. Gallagher},
  \bibinfo{person}{R. Blanco}, {and} \bibinfo{person}{J.~S. Culpepper}.}
  \bibinfo{year}{2017}\natexlab{}.
\newblock \showarticletitle{Efficient cost-aware cascade ranking in multi-stage
  retrieval}. In \bibinfo{booktitle}{{\em Proc. ACM Conf. on Research and
  Development in Information Retrieval (SIGIR)}}. \bibinfo{pages}{445--454}.
\newblock


\bibitem[\protect\citeauthoryear{Choudhury, Culpepper, Bao, and
  Sellis}{Choudhury et~al\mbox{.}}{2018}]%
        {ccbs18-tsas}
\bibfield{author}{\bibinfo{person}{F.~M. Choudhury}, \bibinfo{person}{J.~S.
  Culpepper}, \bibinfo{person}{Z. Bao}, {and} \bibinfo{person}{T. Sellis}.}
  \bibinfo{year}{2018}\natexlab{}.
\newblock \showarticletitle{Batch processing of top-$k$ spatial-textual
  queries}.
\newblock \bibinfo{journal}{{\em ACM Trans. on Spatial Algorithms and
  Systems\/}} \bibinfo{volume}{3}, \bibinfo{number}{4} (\bibinfo{year}{2018}),
  \bibinfo{pages}{1--40}.
\newblock


\bibitem[\protect\citeauthoryear{Cormack, Clarke, and Buettcher}{Cormack
  et~al\mbox{.}}{2009}]%
        {cormack2009reciprocal}
\bibfield{author}{\bibinfo{person}{G.~V. Cormack}, \bibinfo{person}{C.~L.~A.
  Clarke}, {and} \bibinfo{person}{S. Buettcher}.}
  \bibinfo{year}{2009}\natexlab{}.
\newblock \showarticletitle{Reciprocal rank fusion outperforms Condorcet and
  individual rank learning methods}. In \bibinfo{booktitle}{{\em Proc. ACM
  Conf. on Research and Development in Information Retrieval (SIGIR)}}.
  \bibinfo{pages}{758--759}.
\newblock


\bibitem[\protect\citeauthoryear{Crane, Culpepper, Lin, Mackenzie, and
  Trotman}{Crane et~al\mbox{.}}{2017}]%
        {cc+17-wsdm}
\bibfield{author}{\bibinfo{person}{M. Crane}, \bibinfo{person}{J.~S.
  Culpepper}, \bibinfo{person}{J. Lin}, \bibinfo{person}{J. Mackenzie}, {and}
  \bibinfo{person}{A. Trotman}.} \bibinfo{year}{2017}\natexlab{}.
\newblock \showarticletitle{A comparison of Document-at-a-Time and
  Score-at-a-Time query evaluation}. In \bibinfo{booktitle}{{\em Proc. Conf. on
  Web Search and Data Mining (WSDM)}}. \bibinfo{pages}{201--210}.
\newblock


\bibitem[\protect\citeauthoryear{Craswell and Szummer}{Craswell and
  Szummer}{2007}]%
        {cs07-sigir}
\bibfield{author}{\bibinfo{person}{N. Craswell} {and} \bibinfo{person}{M.
  Szummer}.} \bibinfo{year}{2007}\natexlab{}.
\newblock \showarticletitle{Random walks on the click graph}. In
  \bibinfo{booktitle}{{\em Proc. ACM Conf. on Research and Development in
  Information Retrieval (SIGIR)}}. \bibinfo{pages}{239--246}.
\newblock


\bibitem[\protect\citeauthoryear{Dang and Croft}{Dang and Croft}{2010}]%
        {dang2010query}
\bibfield{author}{\bibinfo{person}{V. Dang} {and} \bibinfo{person}{W.~B.
  Croft}.} \bibinfo{year}{2010}\natexlab{}.
\newblock \showarticletitle{Query reformulation using anchor text}. In
  \bibinfo{booktitle}{{\em Proc. Conf. on Web Search and Data Mining (WSDM)}}.
  \bibinfo{pages}{41--50}.
\newblock


\bibitem[\protect\citeauthoryear{Dean and Barroso}{Dean and Barroso}{2013}]%
        {db13-cacm}
\bibfield{author}{\bibinfo{person}{J. Dean} {and} \bibinfo{person}{L.~A.
  Barroso}.} \bibinfo{year}{2013}\natexlab{}.
\newblock \showarticletitle{The tail at scale}.
\newblock \bibinfo{journal}{{\em Comm. ACM\/}} \bibinfo{volume}{56},
  \bibinfo{number}{2} (\bibinfo{year}{2013}), \bibinfo{pages}{74--80}.
\newblock


\bibitem[\protect\citeauthoryear{Dhulipala, Kabiljo, Karrer, Ottaviano,
  Pupyrev, and Shalita}{Dhulipala et~al\mbox{.}}{2016}]%
        {dk+16-kdd}
\bibfield{author}{\bibinfo{person}{L. Dhulipala}, \bibinfo{person}{I. Kabiljo},
  \bibinfo{person}{B. Karrer}, \bibinfo{person}{G. Ottaviano},
  \bibinfo{person}{S. Pupyrev}, {and} \bibinfo{person}{A. Shalita}.}
  \bibinfo{year}{2016}\natexlab{}.
\newblock \showarticletitle{Compressing graphs and indexes with recursive graph
  bisection}. In \bibinfo{booktitle}{{\em Proc. Conf. on Knowledge Discovery
  and Data Mining (WSDM)}}. \bibinfo{pages}{1535--1544}.
\newblock


\bibitem[\protect\citeauthoryear{Dimopoulos, Nepomnyachiy, and Suel}{Dimopoulos
  et~al\mbox{.}}{2013}]%
        {dns13-wsdm}
\bibfield{author}{\bibinfo{person}{C. Dimopoulos}, \bibinfo{person}{S.
  Nepomnyachiy}, {and} \bibinfo{person}{T. Suel}.}
  \bibinfo{year}{2013}\natexlab{}.
\newblock \showarticletitle{Optimizing top-$k$ document retrieval strategies
  for Block-Max indexes}. In \bibinfo{booktitle}{{\em Proc. Conf. on Web Search
  and Data Mining (WSDM)}}. \bibinfo{pages}{113--122}.
\newblock


\bibitem[\protect\citeauthoryear{Din{\c{c}}er, Macdonald, and
  Ounis}{Din{\c{c}}er et~al\mbox{.}}{2014}]%
        {dmo14-sigir}
\bibfield{author}{\bibinfo{person}{B.~T. Din{\c{c}}er}, \bibinfo{person}{C.
  Macdonald}, {and} \bibinfo{person}{I. Ounis}.}
  \bibinfo{year}{2014}\natexlab{}.
\newblock \showarticletitle{Hypothesis testing for the risk-sensitive
  evaluation of retrieval systems}. In \bibinfo{booktitle}{{\em Proc. ACM Conf.
  on Research and Development in Information Retrieval (SIGIR)}}.
  \bibinfo{pages}{23--32}.
\newblock


\bibitem[\protect\citeauthoryear{Ding, Attenberg, Baeza-Yates, and Suel}{Ding
  et~al\mbox{.}}{2011}]%
        {dabs11-wsdm}
\bibfield{author}{\bibinfo{person}{S. Ding}, \bibinfo{person}{J. Attenberg},
  \bibinfo{person}{R. Baeza-Yates}, {and} \bibinfo{person}{T. Suel}.}
  \bibinfo{year}{2011}\natexlab{}.
\newblock \showarticletitle{Batch query processing for web search engines}. In
  \bibinfo{booktitle}{{\em Proc. Conf. on Web Search and Data Mining (WSDM)}}.
  \bibinfo{pages}{137--146}.
\newblock


\bibitem[\protect\citeauthoryear{Ding and Suel}{Ding and Suel}{2011}]%
        {dt11-sigir}
\bibfield{author}{\bibinfo{person}{S. Ding} {and} \bibinfo{person}{T. Suel}.}
  \bibinfo{year}{2011}\natexlab{}.
\newblock \showarticletitle{Faster top-$k$ document retrieval using Block-Max
  indexes}. In \bibinfo{booktitle}{{\em Proc. ACM Conf. on Research and
  Development in Information Retrieval (SIGIR)}}. \bibinfo{pages}{993--1002}.
\newblock


\bibitem[\protect\citeauthoryear{Fagni, Perego, Silvestri, and Orlando}{Fagni
  et~al\mbox{.}}{2006}]%
        {fagni2006boosting}
\bibfield{author}{\bibinfo{person}{T. Fagni}, \bibinfo{person}{R. Perego},
  \bibinfo{person}{F. Silvestri}, {and} \bibinfo{person}{S. Orlando}.}
  \bibinfo{year}{2006}\natexlab{}.
\newblock \showarticletitle{Boosting the performance of web search engines:
  Caching and prefetching query results by exploiting historical usage data}.
\newblock \bibinfo{journal}{{\em ACM Trans. on Information Systems\/}}
  \bibinfo{volume}{24}, \bibinfo{number}{1} (\bibinfo{year}{2006}),
  \bibinfo{pages}{51--78}.
\newblock


\bibitem[\protect\citeauthoryear{Fontoura, Josifovski, Liu, Venkatesan, Zhu,
  and Zien}{Fontoura et~al\mbox{.}}{2011}]%
        {fj+11-pvldb}
\bibfield{author}{\bibinfo{person}{M. Fontoura}, \bibinfo{person}{V.
  Josifovski}, \bibinfo{person}{J. Liu}, \bibinfo{person}{S. Venkatesan},
  \bibinfo{person}{X. Zhu}, {and} \bibinfo{person}{J. Zien}.}
  \bibinfo{year}{2011}\natexlab{}.
\newblock \showarticletitle{Evaluation strategies for top-$k$ queries over
  memory-resident inverted indexes}.
\newblock \bibinfo{journal}{{\em Proc. Conf. on Very Large Databases (VLDB)\/}}
  \bibinfo{volume}{4}, \bibinfo{number}{12} (\bibinfo{year}{2011}),
  \bibinfo{pages}{1213--1224}.
\newblock


\bibitem[\protect\citeauthoryear{Fox and Shaw}{Fox and Shaw}{1994}]%
        {fox1994combination}
\bibfield{author}{\bibinfo{person}{E.~A. Fox} {and} \bibinfo{person}{J.~A.
  Shaw}.} \bibinfo{year}{1994}\natexlab{}.
\newblock \showarticletitle{Combination of multiple searches}.
\newblock \bibinfo{journal}{{\em Proc. Text Retrieval Conf. (TREC)\/}}
  (\bibinfo{year}{1994}), \bibinfo{pages}{243--252}.
\newblock


\bibitem[\protect\citeauthoryear{Fuhr}{Fuhr}{2018}]%
        {fuhr2018some}
\bibfield{author}{\bibinfo{person}{N. Fuhr}.} \bibinfo{year}{2018}\natexlab{}.
\newblock \showarticletitle{Some common mistakes in IR evaluation, and how they
  can be avoided}. In \bibinfo{booktitle}{{\em Proc. ACM Conf. on Research and
  Development in Information Retrieval (SIGIR)}}. \bibinfo{pages}{32--41}.
\newblock


\bibitem[\protect\citeauthoryear{Gan and Suel}{Gan and Suel}{2009}]%
        {gt09-www}
\bibfield{author}{\bibinfo{person}{Q. Gan} {and} \bibinfo{person}{T. Suel}.}
  \bibinfo{year}{2009}\natexlab{}.
\newblock \showarticletitle{Improved techniques for result caching in web
  search engines}. In \bibinfo{booktitle}{{\em Proc. Conf. on the World Wide
  Web (WWW)}}. \bibinfo{pages}{431--440}.
\newblock


\bibitem[\protect\citeauthoryear{Glover, Lawrence, Birmingham, and
  Giles}{Glover et~al\mbox{.}}{1999}]%
        {glover1999architecture}
\bibfield{author}{\bibinfo{person}{E.~J. Glover}, \bibinfo{person}{S.
  Lawrence}, \bibinfo{person}{W.~P. Birmingham}, {and} \bibinfo{person}{C.~L.
  Giles}.} \bibinfo{year}{1999}\natexlab{}.
\newblock \showarticletitle{Architecture of a metasearch engine that supports
  user information needs}. In \bibinfo{booktitle}{{\em Proc. ACM International
  Conf. on Information and Knowledge Management (CIKM)}}.
  \bibinfo{pages}{210--216}.
\newblock


\bibitem[\protect\citeauthoryear{He, Tang, Ouyang, Kang, Yin, and Chang}{He
  et~al\mbox{.}}{2016}]%
        {ht+16-cikm}
\bibfield{author}{\bibinfo{person}{Y. He}, \bibinfo{person}{J. Tang},
  \bibinfo{person}{H. Ouyang}, \bibinfo{person}{C. Kang}, \bibinfo{person}{D.
  Yin}, {and} \bibinfo{person}{Y. Chang}.} \bibinfo{year}{2016}\natexlab{}.
\newblock \showarticletitle{Learning to rewrite queries}. In
  \bibinfo{booktitle}{{\em Proc. ACM International Conf. on Information and
  Knowledge Management (CIKM)}}. \bibinfo{pages}{1443--1452}.
\newblock


\bibitem[\protect\citeauthoryear{Huo, Zhang, Liu, and Ma}{Huo
  et~al\mbox{.}}{2014}]%
        {hzlm14cikm}
\bibfield{author}{\bibinfo{person}{S. Huo}, \bibinfo{person}{M. Zhang},
  \bibinfo{person}{Y. Liu}, {and} \bibinfo{person}{S. Ma}.}
  \bibinfo{year}{2014}\natexlab{}.
\newblock \showarticletitle{Improving tail query performance by fusion model}.
  In \bibinfo{booktitle}{{\em Proc. ACM International Conf. on Information and
  Knowledge Management (CIKM)}}. \bibinfo{pages}{559--658}.
\newblock


\bibitem[\protect\citeauthoryear{J\"arvelin and Kek\"al\"ainen}{J\"arvelin and
  Kek\"al\"ainen}{2002}]%
        {jk02acmtois}
\bibfield{author}{\bibinfo{person}{K. J\"arvelin} {and} \bibinfo{person}{J.
  Kek\"al\"ainen}.} \bibinfo{year}{2002}\natexlab{}.
\newblock \showarticletitle{Cumulated gain-based evaluation of IR techniques}.
\newblock \bibinfo{journal}{{\em ACM Trans. on Information Systems\/}}
  \bibinfo{volume}{20}, \bibinfo{number}{4} (\bibinfo{year}{2002}),
  \bibinfo{pages}{422--446}.
\newblock


\bibitem[\protect\citeauthoryear{Kim, He, Hwang, Elnikety, and Choi}{Kim
  et~al\mbox{.}}{2015}]%
        {sk15-wsdm}
\bibfield{author}{\bibinfo{person}{S. Kim}, \bibinfo{person}{Y. He},
  \bibinfo{person}{S-W. Hwang}, \bibinfo{person}{S. Elnikety}, {and}
  \bibinfo{person}{S. Choi}.} \bibinfo{year}{2015}\natexlab{}.
\newblock \showarticletitle{Delayed-dynamic-selective (DDS) prediction for
  reducing extreme tail latency in web search}. In \bibinfo{booktitle}{{\em
  Proc. Conf. on Web Search and Data Mining (WSDM)}}. \bibinfo{pages}{7--16}.
\newblock


\bibitem[\protect\citeauthoryear{Kong, Scott, and Goerg}{Kong
  et~al\mbox{.}}{2016}]%
        {kjs-16-google}
\bibfield{author}{\bibinfo{person}{J. Kong}, \bibinfo{person}{A. Scott}, {and}
  \bibinfo{person}{G.~M. Goerg}.} \bibinfo{year}{2016}\natexlab{}.
\newblock \showarticletitle{Improving semantic topic clustering for search
  queries with word co-occurrence and bigraph co-clustering}.
\newblock \bibinfo{journal}{{\em Google Inc\/}} (\bibinfo{year}{2016}).
\newblock


\bibitem[\protect\citeauthoryear{Kozorovitsky and Kurland}{Kozorovitsky and
  Kurland}{2011}]%
        {kk11sigir}
\bibfield{author}{\bibinfo{person}{A.~K. Kozorovitsky} {and}
  \bibinfo{person}{O. Kurland}.} \bibinfo{year}{2011}\natexlab{}.
\newblock \showarticletitle{Cluster-based fusion of retrieved lists}. In
  \bibinfo{booktitle}{{\em Proc. ACM Conf. on Research and Development in
  Information Retrieval (SIGIR)}}. \bibinfo{pages}{893--902}.
\newblock


\bibitem[\protect\citeauthoryear{Kurland and Culpepper}{Kurland and
  Culpepper}{2018}]%
        {oc18-tutorial}
\bibfield{author}{\bibinfo{person}{O. Kurland} {and} \bibinfo{person}{J.~S.
  Culpepper}.} \bibinfo{year}{2018}\natexlab{}.
\newblock \showarticletitle{Tutorial / Fusion in information retrieval}. In
  \bibinfo{booktitle}{{\em Proc. ACM Conf. on Research and Development in
  Information Retrieval (SIGIR)}}. \bibinfo{pages}{1383--1386}.
\newblock


\bibitem[\protect\citeauthoryear{Lee, Ai, Croft, and Sheldon}{Lee
  et~al\mbox{.}}{2015}]%
        {lee2015optimization}
\bibfield{author}{\bibinfo{person}{C.-J. Lee}, \bibinfo{person}{Q. Ai},
  \bibinfo{person}{W.~B. Croft}, {and} \bibinfo{person}{D. Sheldon}.}
  \bibinfo{year}{2015}\natexlab{}.
\newblock \showarticletitle{An optimization framework for merging multiple
  result lists}. In \bibinfo{booktitle}{{\em Proc. ACM International Conf. on
  Information and Knowledge Management (CIKM)}}. \bibinfo{pages}{303--312}.
\newblock


\bibitem[\protect\citeauthoryear{Liang, Ren, and de~Rijke}{Liang
  et~al\mbox{.}}{2014}]%
        {lrr14sigir}
\bibfield{author}{\bibinfo{person}{S. Liang}, \bibinfo{person}{Z. Ren}, {and}
  \bibinfo{person}{M. de Rijke}.} \bibinfo{year}{2014}\natexlab{}.
\newblock \showarticletitle{Fusion helps diversification}. In
  \bibinfo{booktitle}{{\em Proc. ACM Conf. on Research and Development in
  Information Retrieval (SIGIR)}}. \bibinfo{pages}{303--312}.
\newblock


\bibitem[\protect\citeauthoryear{Liu}{Liu}{2009}]%
        {l09-ltr}
\bibfield{author}{\bibinfo{person}{T.-Y. Liu}.}
  \bibinfo{year}{2009}\natexlab{}.
\newblock \showarticletitle{Learning to rank for information retrieval}.
\newblock \bibinfo{journal}{{\em Foundations \& Trends in Information
  Retrieval\/}} \bibinfo{volume}{3}, \bibinfo{number}{3}
  (\bibinfo{year}{2009}), \bibinfo{pages}{225--331}.
\newblock


\bibitem[\protect\citeauthoryear{Lu, Moffat, and Culpepper}{Lu
  et~al\mbox{.}}{2016a}]%
        {lmc16irj}
\bibfield{author}{\bibinfo{person}{X. Lu}, \bibinfo{person}{A. Moffat}, {and}
  \bibinfo{person}{J.~S. Culpepper}.} \bibinfo{year}{2016}\natexlab{a}.
\newblock \showarticletitle{The effect of pooling and evaluation depth on IR
  metrics}.
\newblock \bibinfo{journal}{{\em Information Retrieval\/}}
  \bibinfo{volume}{19}, \bibinfo{number}{4} (\bibinfo{year}{2016}),
  \bibinfo{pages}{416--445}.
\newblock


\bibitem[\protect\citeauthoryear{Lu, Moffat, and Culpepper}{Lu
  et~al\mbox{.}}{2016b}]%
        {lmc16ictir}
\bibfield{author}{\bibinfo{person}{X. Lu}, \bibinfo{person}{A. Moffat}, {and}
  \bibinfo{person}{J.~S. Culpepper}.} \bibinfo{year}{2016}\natexlab{b}.
\newblock \showarticletitle{Efficient and effective higher order proximity
  modeling}. In \bibinfo{booktitle}{{\em Proc. International Conf. on Theory of
  Information Retrieval (ICTIR)}}. \bibinfo{pages}{21--30}.
\newblock


\bibitem[\protect\citeauthoryear{Ma and Wang}{Ma and Wang}{2012}]%
        {mw12-sigir}
\bibfield{author}{\bibinfo{person}{H. Ma} {and} \bibinfo{person}{B. Wang}.}
  \bibinfo{year}{2012}\natexlab{}.
\newblock \showarticletitle{User-aware caching and prefetching query results in
  web search engines}. In \bibinfo{booktitle}{{\em Proc. ACM Conf. on Research
  and Development in Information Retrieval (SIGIR)}}.
  \bibinfo{pages}{1163--1164}.
\newblock


\bibitem[\protect\citeauthoryear{Macdonald, Santos, and Ounis}{Macdonald
  et~al\mbox{.}}{2013a}]%
        {mso13irj}
\bibfield{author}{\bibinfo{person}{C. Macdonald}, \bibinfo{person}{R.~L.~T.
  Santos}, {and} \bibinfo{person}{I. Ounis}.} \bibinfo{year}{2013}\natexlab{a}.
\newblock \showarticletitle{The whens and hows of learning to rank for web
  search}.
\newblock \bibinfo{journal}{{\em Information Retrieval\/}}
  \bibinfo{volume}{16}, \bibinfo{number}{5} (\bibinfo{year}{2013}),
  \bibinfo{pages}{584--628}.
\newblock


\bibitem[\protect\citeauthoryear{Macdonald, Santos, Ounis, and He}{Macdonald
  et~al\mbox{.}}{2013b}]%
        {msoh13acmtois}
\bibfield{author}{\bibinfo{person}{C. Macdonald}, \bibinfo{person}{R.~L.~T.
  Santos}, \bibinfo{person}{I. Ounis}, {and} \bibinfo{person}{B. He}.}
  \bibinfo{year}{2013}\natexlab{b}.
\newblock \showarticletitle{About learning models with multiple query-dependent
  features}.
\newblock \bibinfo{journal}{{\em ACM Trans. on Information Systems\/}}
  \bibinfo{volume}{31}, \bibinfo{number}{3} (\bibinfo{year}{2013}),
  \bibinfo{pages}{11:1--11:39}.
\newblock


\bibitem[\protect\citeauthoryear{Mackenzie}{Mackenzie}{2017}]%
        {m17-sigir}
\bibfield{author}{\bibinfo{person}{J. Mackenzie}.}
  \bibinfo{year}{2017}\natexlab{}.
\newblock \showarticletitle{Managing tail latencies in large scale IR systems}.
  In \bibinfo{booktitle}{{\em Proc. ACM Conf. on Research and Development in
  Information Retrieval (SIGIR)}}. \bibinfo{pages}{1369}.
\newblock


\bibitem[\protect\citeauthoryear{Mackenzie, Culpepper, Blanco, Crane, and
  Lin}{Mackenzie et~al\mbox{.}}{2018}]%
        {mcbcl18-wsdm}
\bibfield{author}{\bibinfo{person}{J. Mackenzie}, \bibinfo{person}{J.~S.
  Culpepper}, \bibinfo{person}{R. Blanco}, \bibinfo{person}{M. Crane}, {and}
  \bibinfo{person}{J. Lin}.} \bibinfo{year}{2018}\natexlab{}.
\newblock \showarticletitle{Query driven algorithm selection in early stage
  retrieval}. In \bibinfo{booktitle}{{\em Proc. Conf. on Web Search and Data
  Mining (WSDM)}}. \bibinfo{pages}{396--404}.
\newblock


\bibitem[\protect\citeauthoryear{Mallia, Ottaviano, Porciani, Tonellotto, and
  Venturini}{Mallia et~al\mbox{.}}{2017}]%
        {mo+17-sigir}
\bibfield{author}{\bibinfo{person}{A. Mallia}, \bibinfo{person}{G. Ottaviano},
  \bibinfo{person}{E. Porciani}, \bibinfo{person}{N. Tonellotto}, {and}
  \bibinfo{person}{R. Venturini}.} \bibinfo{year}{2017}\natexlab{}.
\newblock \showarticletitle{Faster BlockMax WAND with variable-sized blocks}.
  In \bibinfo{booktitle}{{\em Proc. ACM Conf. on Research and Development in
  Information Retrieval (SIGIR)}}. \bibinfo{pages}{625--634}.
\newblock


\bibitem[\protect\citeauthoryear{Metzler and Croft}{Metzler and Croft}{2005}]%
        {mc05-sigir}
\bibfield{author}{\bibinfo{person}{D. Metzler} {and} \bibinfo{person}{W.~B.
  Croft}.} \bibinfo{year}{2005}\natexlab{}.
\newblock \showarticletitle{A Markov random field model for term dependencies}.
  In \bibinfo{booktitle}{{\em Proc. ACM Conf. on Research and Development in
  Information Retrieval (SIGIR)}}. \bibinfo{pages}{472--479}.
\newblock


\bibitem[\protect\citeauthoryear{Metzler, Dumais, and Meek}{Metzler
  et~al\mbox{.}}{2007}]%
        {metzler2007similarity}
\bibfield{author}{\bibinfo{person}{D. Metzler}, \bibinfo{person}{S. Dumais},
  {and} \bibinfo{person}{C. Meek}.} \bibinfo{year}{2007}\natexlab{}.
\newblock \showarticletitle{Similarity measures for short segments of text}. In
  \bibinfo{booktitle}{{\em Proc. European Conf. in Information Retrieval
  (ECIR)}}. \bibinfo{pages}{16--27}.
\newblock


\bibitem[\protect\citeauthoryear{Moffat}{Moffat}{2016}]%
        {moffat16adcs}
\bibfield{author}{\bibinfo{person}{A. Moffat}.}
  \bibinfo{year}{2016}\natexlab{}.
\newblock \showarticletitle{Judgment pool effects caused by query variations}.
  In \bibinfo{booktitle}{{\em Proc. Australasian Document Computing Symp.
  (ADCS)}}. \bibinfo{pages}{65--68}.
\newblock


\bibitem[\protect\citeauthoryear{Moffat, Bailey, Scholer, and Thomas}{Moffat
  et~al\mbox{.}}{2017}]%
        {mbst17acmtois}
\bibfield{author}{\bibinfo{person}{A. Moffat}, \bibinfo{person}{P. Bailey},
  \bibinfo{person}{F. Scholer}, {and} \bibinfo{person}{P. Thomas}.}
  \bibinfo{year}{2017}\natexlab{}.
\newblock \showarticletitle{Incorporating user expectations and behavior into
  the measurement of search effectiveness}.
\newblock \bibinfo{journal}{{\em ACM Trans. on Information Systems\/}}
  \bibinfo{volume}{35}, \bibinfo{number}{3} (\bibinfo{year}{2017}),
  \bibinfo{pages}{24:1--24:38}.
\newblock


\bibitem[\protect\citeauthoryear{Moffat and Zobel}{Moffat and Zobel}{2008}]%
        {mz08acmtois}
\bibfield{author}{\bibinfo{person}{A. Moffat} {and} \bibinfo{person}{J.
  Zobel}.} \bibinfo{year}{2008}\natexlab{}.
\newblock \showarticletitle{Rank-biased precision for measurement of retrieval
  effectiveness}.
\newblock \bibinfo{journal}{{\em ACM Trans. on Information Systems\/}}
  \bibinfo{volume}{27}, \bibinfo{number}{1} (\bibinfo{year}{2008}),
  \bibinfo{pages}{2.1--2.27}.
\newblock


\bibitem[\protect\citeauthoryear{Mour{\~a}o and Magalh{\~a}es}{Mour{\~a}o and
  Magalh{\~a}es}{2018}]%
        {mm18-cikm}
\bibfield{author}{\bibinfo{person}{A. Mour{\~a}o} {and} \bibinfo{person}{J.
  Magalh{\~a}es}.} \bibinfo{year}{2018}\natexlab{}.
\newblock \showarticletitle{Low-complexity supervised rank fusion models}. In
  \bibinfo{booktitle}{{\em Proc. ACM International Conf. on Information and
  Knowledge Management (CIKM)}}. \bibinfo{pages}{1691--1694}.
\newblock


\bibitem[\protect\citeauthoryear{Ottaviano and Venturini}{Ottaviano and
  Venturini}{2014}]%
        {go14-sigir}
\bibfield{author}{\bibinfo{person}{G. Ottaviano} {and} \bibinfo{person}{R.
  Venturini}.} \bibinfo{year}{2014}\natexlab{}.
\newblock \showarticletitle{Partitioned Elias-Fano indexes}. In
  \bibinfo{booktitle}{{\em Proc. ACM Conf. on Research and Development in
  Information Retrieval (SIGIR)}}. \bibinfo{pages}{273--282}.
\newblock


\bibitem[\protect\citeauthoryear{Radlinski, Kurup, and Joachims}{Radlinski
  et~al\mbox{.}}{2008}]%
        {radlinski2008does}
\bibfield{author}{\bibinfo{person}{F. Radlinski}, \bibinfo{person}{M. Kurup},
  {and} \bibinfo{person}{T. Joachims}.} \bibinfo{year}{2008}\natexlab{}.
\newblock \showarticletitle{How does clickthrough data reflect retrieval
  quality?}. In \bibinfo{booktitle}{{\em Proc. ACM International Conf. on
  Information and Knowledge Management (CIKM)}}. \bibinfo{pages}{43--52}.
\newblock


\bibitem[\protect\citeauthoryear{Scholer and Williams}{Scholer and
  Williams}{2002}]%
        {scholer2002query}
\bibfield{author}{\bibinfo{person}{F. Scholer} {and} \bibinfo{person}{H.~E.
  Williams}.} \bibinfo{year}{2002}\natexlab{}.
\newblock \showarticletitle{Query association for effective retrieval}. In
  \bibinfo{booktitle}{{\em Proc. ACM International Conf. on Information and
  Knowledge Management (CIKM)}}. \bibinfo{pages}{324--331}.
\newblock


\bibitem[\protect\citeauthoryear{Sheldon, Shokouhi, Szummer, and
  Craswell}{Sheldon et~al\mbox{.}}{2011}]%
        {sssc11wsdm}
\bibfield{author}{\bibinfo{person}{D. Sheldon}, \bibinfo{person}{M. Shokouhi},
  \bibinfo{person}{M. Szummer}, {and} \bibinfo{person}{N. Craswell}.}
  \bibinfo{year}{2011}\natexlab{}.
\newblock \showarticletitle{LambdaMerge: Merging the results of query
  reformulations}. In \bibinfo{booktitle}{{\em Proc. Conf. on Web Search and
  Data Mining (WSDM)}}. \bibinfo{pages}{795--804}.
\newblock


\bibitem[\protect\citeauthoryear{Shen, Karimzadehgan, Bendersky, Qin, and
  Metzler}{Shen et~al\mbox{.}}{2018}]%
        {sk+18-cikm}
\bibfield{author}{\bibinfo{person}{J. Shen}, \bibinfo{person}{M.
  Karimzadehgan}, \bibinfo{person}{M. Bendersky}, \bibinfo{person}{Z. Qin},
  {and} \bibinfo{person}{D. Metzler}.} \bibinfo{year}{2018}\natexlab{}.
\newblock \showarticletitle{Multi-task learning for email search ranking with
  auxiliary query clustering}. In \bibinfo{booktitle}{{\em Proc. ACM
  International Conf. on Information and Knowledge Management (CIKM)}}.
  \bibinfo{pages}{2127--2135}.
\newblock


\bibitem[\protect\citeauthoryear{Strohman, Turtle, and Croft}{Strohman
  et~al\mbox{.}}{2005}]%
        {stc05-sigir}
\bibfield{author}{\bibinfo{person}{T. Strohman}, \bibinfo{person}{H. Turtle},
  {and} \bibinfo{person}{W.~B. Croft}.} \bibinfo{year}{2005}\natexlab{}.
\newblock \showarticletitle{Optimization strategies for complex queries}. In
  \bibinfo{booktitle}{{\em Proc. ACM Conf. on Research and Development in
  Information Retrieval (SIGIR)}}. \bibinfo{pages}{219--225}.
\newblock


\bibitem[\protect\citeauthoryear{Turtle and Flood}{Turtle and Flood}{1995}]%
        {tf95ipm}
\bibfield{author}{\bibinfo{person}{H.~R. Turtle} {and} \bibinfo{person}{J.
  Flood}.} \bibinfo{year}{1995}\natexlab{}.
\newblock \showarticletitle{Query evaluation: Strategies and optimizations}.
\newblock \bibinfo{journal}{{\em Information Processing \& Management\/}}
  \bibinfo{volume}{31}, \bibinfo{number}{6} (\bibinfo{year}{1995}),
  \bibinfo{pages}{831--850}.
\newblock


\bibitem[\protect\citeauthoryear{Vogt}{Vogt}{2000}]%
        {vogt2000much}
\bibfield{author}{\bibinfo{person}{C.~C. Vogt}.}
  \bibinfo{year}{2000}\natexlab{}.
\newblock \showarticletitle{How much more is better? Characterizing the effects
  of adding more IR systems to a combination}. In \bibinfo{booktitle}{{\em
  Proc. Recherche d'Information etses Applications (RIAO)}}.
  \bibinfo{pages}{457--475}.
\newblock


\bibitem[\protect\citeauthoryear{Vogt and Cottrell}{Vogt and Cottrell}{1999a}]%
        {vogt1999adaptive}
\bibfield{author}{\bibinfo{person}{C.~C. Vogt} {and} \bibinfo{person}{G.~W.
  Cottrell}.} \bibinfo{year}{1999}\natexlab{a}.
\newblock \bibinfo{booktitle}{{\em Adaptive combination of evidence for
  information retrieval}}.
\newblock \bibinfo{publisher}{University of California, San Diego}.
\newblock


\bibitem[\protect\citeauthoryear{Vogt and Cottrell}{Vogt and Cottrell}{1999b}]%
        {vc99irj}
\bibfield{author}{\bibinfo{person}{C.~C. Vogt} {and} \bibinfo{person}{G.~W.
  Cottrell}.} \bibinfo{year}{1999}\natexlab{b}.
\newblock \showarticletitle{Fusion via a linear combination of scores}.
\newblock \bibinfo{journal}{{\em Information Retrieval\/}} \bibinfo{volume}{1},
  \bibinfo{number}{3} (\bibinfo{year}{1999}), \bibinfo{pages}{151--173}.
\newblock


\bibitem[\protect\citeauthoryear{Wang, Lo, Yiu, Tong, Wang, and Liu}{Wang
  et~al\mbox{.}}{2014}]%
        {wl+14-tois}
\bibfield{author}{\bibinfo{person}{J. Wang}, \bibinfo{person}{E. Lo},
  \bibinfo{person}{M.~L. Yiu}, \bibinfo{person}{J. Tong}, \bibinfo{person}{G.
  Wang}, {and} \bibinfo{person}{X. Liu}.} \bibinfo{year}{2014}\natexlab{}.
\newblock \showarticletitle{Cache design of SSD-based search engine
  architectures: An experimental study}.
\newblock \bibinfo{journal}{{\em ACM Trans. on Information Systems\/}}
  \bibinfo{volume}{32}, \bibinfo{number}{4} (\bibinfo{year}{2014}),
  \bibinfo{pages}{1--26}.
\newblock


\bibitem[\protect\citeauthoryear{Wen, Nie, and Zhang}{Wen
  et~al\mbox{.}}{2001}]%
        {wen2001clustering}
\bibfield{author}{\bibinfo{person}{J.~R. Wen}, \bibinfo{person}{J.~Y. Nie},
  {and} \bibinfo{person}{H.~J. Zhang}.} \bibinfo{year}{2001}\natexlab{}.
\newblock \showarticletitle{Clustering user queries of a search engine}. In
  \bibinfo{booktitle}{{\em Proc. Conf. on the World Wide Web (WWW)}}.
  \bibinfo{pages}{162--168}.
\newblock


\bibitem[\protect\citeauthoryear{Wen, Nie, and Zhang}{Wen
  et~al\mbox{.}}{2002}]%
        {wen2002query}
\bibfield{author}{\bibinfo{person}{J.~R. Wen}, \bibinfo{person}{J.~Y. Nie},
  {and} \bibinfo{person}{H.~J. Zhang}.} \bibinfo{year}{2002}\natexlab{}.
\newblock \showarticletitle{Query clustering using user logs}.
\newblock \bibinfo{journal}{{\em ACM Trans. on Information Systems\/}}
  \bibinfo{volume}{20}, \bibinfo{number}{1} (\bibinfo{year}{2002}),
  \bibinfo{pages}{59--81}.
\newblock


\bibitem[\protect\citeauthoryear{Xue, Zeng, Chen, Yu, Ma, Xi, and Fan}{Xue
  et~al\mbox{.}}{2004}]%
        {xz+04-cikm}
\bibfield{author}{\bibinfo{person}{G-R. Xue}, \bibinfo{person}{H-J. Zeng},
  \bibinfo{person}{Z. Chen}, \bibinfo{person}{Y. Yu}, \bibinfo{person}{W-Y.
  Ma}, \bibinfo{person}{W. Xi}, {and} \bibinfo{person}{W. Fan}.}
  \bibinfo{year}{2004}\natexlab{}.
\newblock \showarticletitle{Optimizing web search using web click-through
  data}. In \bibinfo{booktitle}{{\em Proc. ACM International Conf. on
  Information and Knowledge Management (CIKM)}}. \bibinfo{pages}{118--126}.
\newblock


\bibitem[\protect\citeauthoryear{Xue and Croft}{Xue and Croft}{2013}]%
        {xc13tois}
\bibfield{author}{\bibinfo{person}{X. Xue} {and} \bibinfo{person}{W.~B.
  Croft}.} \bibinfo{year}{2013}\natexlab{}.
\newblock \showarticletitle{Modeling reformulation using query distributions}.
\newblock \bibinfo{journal}{{\em ACM Trans. on Information Systems\/}}
  \bibinfo{volume}{31}, \bibinfo{number}{2} (\bibinfo{year}{2013}),
  \bibinfo{pages}{6:1--6:34}.
\newblock


\bibitem[\protect\citeauthoryear{Yun, He, Elnikety, and Ren}{Yun
  et~al\mbox{.}}{2015}]%
        {yh+15-sigir}
\bibfield{author}{\bibinfo{person}{J-M. Yun}, \bibinfo{person}{Y. He},
  \bibinfo{person}{S. Elnikety}, {and} \bibinfo{person}{S. Ren}.}
  \bibinfo{year}{2015}\natexlab{}.
\newblock \showarticletitle{Optimal aggregation policy for reducing tail
  latency of web search}. In \bibinfo{booktitle}{{\em Proc. ACM Conf. on
  Research and Development in Information Retrieval (SIGIR)}}.
  \bibinfo{pages}{63--72}.
\newblock


\bibitem[\protect\citeauthoryear{Zobel and Moffat}{Zobel and Moffat}{2006}]%
        {zm06compsurv}
\bibfield{author}{\bibinfo{person}{J. Zobel} {and} \bibinfo{person}{A.
  Moffat}.} \bibinfo{year}{2006}\natexlab{}.
\newblock \showarticletitle{Inverted files for text search engines}.
\newblock \bibinfo{journal}{{\it Comput. Surveys}} \bibinfo{volume}{38},
  \bibinfo{number}{2} (\bibinfo{year}{2006}), \bibinfo{pages}{6.1--6.56}.
\newblock


\end{thebibliography}

\end{document}